\documentclass[preprint,12pt]{elsarticle}

\usepackage{amssymb, mathdots}
\usepackage{color}

\journal{}

\begin{document}

\begin{frontmatter}



\title{An explicit expression for Euclidean self-dual cyclic codes of length $2^k$ over Galois ring ${\rm GR}(4,m)$}


\author{Yuan Cao$^{a, \ b, \ c}$, Yonglin Cao$^{a, \ \ast}$, San Ling$^{d}$, Guidong Wang$^{a}$}

\address{$^{a}$School of Mathematics and Statistics,
Shandong University of Technology, Zibo, Shandong 255091, China

\vskip 1mm $^{b}$Hubei Key Laboratory of Applied Mathematics, Faculty of Mathematics and Statistics, Hubei University, Wuhan 430062, China

\vskip 1mm $^{c}$School of Mathematics and Statistics, Changsha University of Science and Technology, Changsha, Hunan 410114, China

\vskip 1mm $^{d}$School of Physical and Mathematical Sciences, Nanyang Technological University, 21 Nanyang Link, Singapore 637371, Republic of Singapore}
\cortext[cor1]{corresponding author.  \\
E-mail addresses: yuancao@sdut.edu.cn (Yuan Cao), \ ylcao@sdut.edu.cn (Yonglin Cao),
\ lingsan@ntu.edu.sg (S. Ling), \ hbuwgd@163.com (G. Wang).}

\begin{abstract}
For any positive integers $m$ and $k$,
existing literature only determines the number of
all Euclidean self-dual cyclic codes of length $2^k$
over the Galois ring ${\rm GR}(4,m)$, such as in [Des. Codes Cryptogr. (2012) 63:105--112].
Using properties for Kronecker products of matrices of a specific type and column vectors of these matrices, we give a simple and efficient method to construct all
these self-dual cyclic codes precisely. On this basis,
we provide an explicit expression to accurately represent all distinct Euclidean self-dual cyclic codes of length $2^k$ over ${\rm GR}(4,m)$, using combination numbers.
As an application, we list all distinct Euclidean self-dual cyclic codes over ${\rm GR}(4,m)$ of length $2^k$ explicitly,
for $k=4,5,6$.
\end{abstract}

\begin{keyword}
Cyclic code; Euclidean self-dual code; Galois ring; Combination number; Kronecker product of matrices


\vskip 3mm
\noindent
{\small {\bf Mathematics Subject Classification (2000)} \  94B15, 94B05, 11T71}
\end{keyword}

\end{frontmatter}


\section{Introduction}
\noindent
   The construction of self-dual codes is an interesting topic in coding theory, due to that this class of codes is closely related to
other fields of mathematics such as lattices, cryptography, invariant theory, block designs, etc.
The study of cyclic codes over finite rings started in the 1990s. It was motivated by the discovery that some good
non-linear codes over $\mathbb{Z}_2$ can be viewed as binary images under a Gray
map of linear cyclic codes over $\mathbb{Z}_4$ \cite{s8}.
Importantly, self-dual codes over $\mathbb{Z}_4$ relate to combinatorial designs
and unimodular lattices (cf. \cite{Bachoc97, Bonnecaze00,Calderbank1997,Gaborit2006,Harada2002,Harada2005,Harada2011,Harada1999}).

\par
  Cyclic codes were initially studied where their lengths are relatively prime to the characteristic of the ring since this condition simplifies the algebraic structure of the code significantly. Subsequently, cyclic codes whose lengths are not relatively prime to the characteristic of the ring were also studied.
 The first step was done by
  Abualrub and Oehmke in \cite{s1}, where they  determined the generators for cyclic codes over $\mathbb{Z}_4$ for lengths of the form $2^k$.

\par
   In general,
let $p$ be an arbitrary prime number and let $e, N$ be positive integers.
Cyclic codes of length $N$ over the ring $\mathbb{Z}_{p^e}=\{0,1,\ldots,p^e-1\}$ can be viewed as ideals
with in the ring $\frac{\mathbb{Z}_{p^e}[x]}{\langle x^N-1\rangle}$. Namely, any vector
$(c_0,c_1,\ldots,c_{N-1})$ is associated with the polynomial $c_0+c_1x+\ldots+c_{N-1}x^{N-1}$.
Let
$N=p^kn$ where ${\rm gcd}(p,n)=1$ and $k\geq 1$. The ring $\frac{\mathbb{Z}_{p^e}[x]}{\langle x^N-1\rangle}$
is isomorphic to the direct
sum of rings of the form $\frac{{\rm GR}(p^e,m)[x]}{\langle x^{p^k}-1\rangle}$ (cf. \cite{s2,s4}), where ${\rm GR}(p^e,m)$ denotes the Galois ring
of characteristic $p^e$  with $p^{em}$ elements \cite{s10}.

\par
 When $e=2$,
using some known
results and the standard Discrete Fourier Transform decomposition, Jitman {\it et al.} \cite{Jitman}
studied cyclic codes of
length $p^kn$ over ${\rm GR}(p^2,s)$:
  Let $C$ be a cyclic code of length $p^kn$ over ${\rm GR}(p^2,s)$. By \cite[Lemma 4.3]{Jitman}, we know that
$C\cong \prod_{i\in \mathcal{J}_0}C_i\times \prod_{j\in \mathcal{J}_1}C_j\times \prod_{k\in \mathcal{J}_2}C_k,$
where

\par
  $C_t$ is a cyclic code of length $p^k$ over the Galois ring ${\rm GR}(p^2,sm_t)$,

 for every $t\in \mathcal{J}_0\cup \mathcal{J}_1\cup \mathcal{J}_2$ (please refer to \cite{Jitman}).

\noindent
   In particular, the cyclic code $C$ is Euclidean self-dual if and only if the following three conditions are satisfied
(See \cite[Proposition 4.5]{Jitman}) :

\par
  $\diamond$ $C_i$ is a Euclidean self-dual code over ${\rm GR}(p^2,sm_i)$, for all
$i\in \mathcal{J}_0$;

\par
  $\diamond$ $C_j$ is a Hermitian self-dual code over ${\rm GR}(p^2,sm_j)$, for all $j\in \mathcal{J}_1$;

\par
  $\diamond$ $C_k =C^{\bot_E}_{n-k \ ({\rm mod} \ n)}$
for all $k\in \mathcal{J}_2$.

\noindent
So, in order to construct Euclidean self-dual cyclic codes of length $p^kn$ over ${\rm GR}(p^2,s)$ efficiently and explicitly,
the first thing need to do is:

\noindent
 \textsf{Give an explicit representation and an efficient construction method for all distinct Euclidean (Hermitian) self-dual cyclic codes
of length $p^k$ over the Galois ring ${\rm GR}(p^2,m)$, for
arbitrary positive integers $k$ and $m$}.

\par
  This paper focuses on Euclidean self-dual cyclic codes of length $2^k$ over ${\rm GR}(4,m)$.
  Main works on this problem have progressed as follows:

\par
   $\flat$)
   Dougherty and Ling \cite{s5} gave a representation for all cyclic codes of length $2^k$ over ${\rm GR}(4,m)$.
Modifying their approach,  Kiah \textit{et al.} \cite{s6}
determined all cyclic codes of length $p^k$ over ${\rm GR}(p^2,m)$, where $p$ is odd.

\par
   $\natural$) For any prime number $p$, Kiah \textit{et al.}
proved that the number of Euclidean self-dual codes of
length $p^k$ over ${\rm GR}(p^2,m)$ is equal to the number of solutions of certain matrix equations in \cite{s6}.

\par
  A year later, using a similar approach, Sobhani and Esmaeili \cite{s9} found an error for the matrix equations in \cite{s6} when $p=2$,
Then they corrected the error and provided the correct matrix equations.

  \textsf{However, the number of solutions and solutions of these matrix equations had not been determined
in \cite{s6} and \cite{s9}}.

\par
   $\sharp$)
  With the aid of Genocchi numbers, Kiah \textit{et al.} \cite{s7} determined the number
of solutions for the matrix equations provided by \cite{s9}. Based on this,
the number of Euclidean self-dual cyclic codes of length $2^k$ over ${\rm GR}(4,m)$ was given by:
$N_{{\rm E}}({\rm GR}(4,m),2^k)=1+2^m+2(2^m)^2\cdot\frac{(2^m)^{2^{k-2}-1}-1}{2^m-1}$ (see \cite[Corollary 3.5]{s7}).
\textsf{But explicit solutions of the matrix equations were not given}.

\par
   To the best of our knowledge,
 the following question remain unresolved:

\noindent
  \textsf{How to construct all distinct Euclidean self-dual cyclic codes
 of length $2^k$ over ${\rm GR}(4,m)$ precisely and efficiently? for arbitrary positive integers $k$ and $m$.}

\par
  Recently,  Cao \textit{et al.}  \cite{s3} gave an efficient method for the construction of all distinct Euclidean self-dual cyclic codes of length $2^s$ over the finite chain ring $\mathbb{F}_{2^m}+u\mathbb{F}_{2^m}$
$(u^2=0)$, by use of properties of the Kronecker product of matrices and calculation for certain types of homogeneous linear equations over $\mathbb{F}_{2^m}$,
where $\mathbb{F}_{2^m}$ is the finite field of $2^m$ elements.
Modifying this approach,
in this paper we provide an
efficient method to construct and express all Euclidean self-dual cyclic codes of length $2^k$ over
the Galois ring ${\rm GR}(4,m)$.

   Here are some necessary concepts and natation: Let $R={\rm GR}(4,m)$ and $R^{2^k}=\{(a_0,a_1,\ldots,a_{2^k-1})\mid a_i\in R\}$. The
standard \textit{Euclidean inner-product} is defined by: $[\underline{a},\underline{b}]=\sum_{i=0}^{2^k-1}a_ib_i\in R$,
for all $\underline{a}=(a_0,a_1,\ldots,a_{2^k-1})$ and $\underline{b}=(b_0,b_1,\ldots,b_{2^k-1})$ with
$a_i,b_i\in R$.
Let $C$ be a linear code of length $2^k$ over $R$. Its
\textit{Euclidean dual code} is defined as $C^{\bot}=\{\underline{a}\in R^{2^k}\mid [\underline{a},\underline{b}]=0, \
\forall \underline{b}\in C\}$. Then $C$ is called a \textit{Euclidean self-dual code} if
$C^{\bot}=C$. Frow now on, all self-dual codes in this paper are Euclidean self-dual codes.

\par
  The paper is organised as follows. In Section 2, we provide an explicit
representation for the solutions of certain homogeneous linear equations over the finite field
$\mathbb{F}_{2^m}$, using Kronecker products of matrices of specific types.
Then by Theorem 2, we give a simple and efficient method to construct all distinct self-dual cyclic codes of length $2^k$ over ${\rm GR}(4,m)$ precisely.
On this basis,
we provide an explicit expression to accurately represent all these self-dual codes by Theorem 3, using combination numbers.
In Section 3, we give a detailed proof of Theorem 2.
In Section 4, we list all distinct Euclidean self-dual cyclic codes over ${\rm GR}(4,m)$ of length $2^k$ explicitly,
for $k=4,5,6$. Section 5 concludes the paper.



\section{Euclidean self-dual cyclic codes of length $2^k$ over ${\rm GR}(4,m)$}
\noindent
In this section, we determine the solution spaces of certain homogeneous linear equations over
$\mathbb{F}_{2^m}$. Then we give a direct approach to construct all
distinct self-dual cyclic codes of length $2^k$ over the Galois ring ${\rm GR}(4,m)$ precisely and
express these self-dual cyclic codes explicitly.

\par
   Let $A=(a_{ij})$ and $B$ be matrices over $\mathbb{F}_{2^m}$ of sizes $s\times t$ and $l\times v$ respectively.
We denote by $A^{{\rm tr}}$ the transpose of $A$.
Recall that the \textit{Kronecker product} of $A$ and $B$ is
defined by $A\otimes B=(a_{ij}B)$ which is a matrix over $\mathbb{F}_{2^m}$ of size $sl\times tv$. Then
we define
\begin{equation}
\label{eq1}
G_2=\left(\begin{array}{cc} 1 & 0 \cr 1 & 1\end{array}\right), \ G_{2^\lambda}=G_2\otimes G_{2^{\lambda-1}}=\left(\begin{array}{cc} G_{2^{\lambda-1}} & 0 \cr G_{2^{\lambda-1}} & G_{2^{\lambda-1}}\end{array}\right)
\ {\rm for} \ {\rm all} \ \lambda\geq 2.
\end{equation}
Moreover, for any integer $l$, $1\leq l\leq 2^\lambda$, let $I_l$ be the identity matrix of order $l$ and denote by
$M_l$ the submatrix of size $l\times l$ in the upper left corner of
$I_{2^\lambda}+G_{2^\lambda}$, i.e.,
\begin{equation}
\label{eq2}
\left(\begin{array}{cc} M_l & 0 \cr \ast & \ast\end{array}\right)=I_{2^\lambda}+G_{2^\lambda},
\end{equation}
where $M_l$ is a matrix over $\mathbb{F}_{2}$ of size $l\times l$.
In particular, we have
$M_{2^\lambda}=I_{2^\lambda}+G_{2^\lambda}.$

\par
  For the purpose of expressing the conclusion of this paper, we label the rows of the matrix $M_l$ from top to bottom as: $0$th row,  $1$st row, $\ldots$, $(l-1)$st row, and label the columns of $M_l$ from left to right as: $1$st column,  $2$nd column, $\ldots$,
  $l$th column.

   The solution space of the homogeneous linear equations with
coefficient matrix $M_l$ is determined by the following
theorem when $l$ is odd.

\vskip 3mm\noindent
 {\bf Theorem 1}
  \textit{Let $s$ be an arbitrary positive integer. We denote by $\Upsilon_{j}^{[0,2s-1)}$ the $j$th column vector
of the matrix $M_{2s-1}$ for all $j=1,2,\ldots,2s-1$, i.e.,
$$M_{2s-1}=\left(\Upsilon_{1}^{[0,2s-1)},\Upsilon_{2}^{[0,2s-1)},\ldots,\Upsilon_{2s-1}^{[0,2s-1)}\right)
\ {\mbox with} \ \Upsilon_{j}^{[0,2s-1)}\in\mathbb{F}_2^{2s-1}.$$
Let $\varepsilon_{2s-1}=(0,\ldots,0,1)^{{\rm tr}}\in\mathbb{F}_2^{2s-1}$ and let
$\mathcal{S}_{2s-1}$ be the solution space for the homogeneous linear equations over $\mathbb{F}_{2^m}$}:
\begin{equation}
\label{eq3}
M_{2s-1}(y_0, y_1, \ldots, y_{2s-2})^{{\rm tr}}=(0,0,\ldots,0)^{{\rm tr}}.
\end{equation}
\textit{Then we have the following conclusions}:
\begin{itemize}
 \item[(i)]
 \textit{${\rm dim}_{\mathbb{F}_{2^m}}(\mathcal{S}_{2s-1})=s$ and the following $s$ column vectors:
$$\Upsilon_{1}^{[0,2s-1)}, \Upsilon_{3}^{[0,2s-1)},\ldots,\Upsilon_{2s-3}^{[0,2s-1)}, \varepsilon_{2s-1}$$
form a basis of the $\mathbb{F}_{2^m}$-linear space $\mathcal{S}_{2s-1}$}.

\item[(ii)]
 $\mathcal{S}_{2s-1}=\{\sum_{i=1}^{s-1}a_{2i-1}\Upsilon_{2i-1}^{[0,2s-1)}+a_{2s-2}\varepsilon_{2s-1}
 \mid a_{2i-1}\in\mathbb{F}_{2^m}
 \ {\rm for} \ {\rm all} \ 1\leq i\leq s-1,  \ {\rm and} \ a_{2s-2}\in\mathbb{F}_{2^m}\}$.
\end{itemize}

\noindent
 {\bf Proof.}
 Obviously, we have $G_2^2=I_2$. Let $\lambda\geq 2$ and assume that $G_{2^{\lambda-1}}^2=I_{2^{\lambda-1}}$.
Then by Eq. (\ref{eq1}), we have $G_{2^\lambda}^2=G_2^2\otimes G_{2^{\lambda-1}}^2=I_2\otimes I_{2^{\lambda-1}}
=I_{2^\lambda}$.
According to the principle of mathematical induction, we conclude that
$G_{2^\lambda}^2=I_{2^\lambda}$ for any positive integer $\lambda$. This implies
$$M_{2^\lambda}^2=(G_{2^\lambda}+I_{2^\lambda})^2=G_{2^\lambda}^2+I_{2^\lambda}=0.$$

\par
  Now, let $\lambda$ be the least positive integer satisfying $2s-1< 2^\lambda$.
By Eq. (\ref{eq2}), it follows that
 $$M_{2^\lambda}^2=(G_{2^\lambda}+I_{2^\lambda})^2=\left(\begin{array}{cc} M_{2s-1} & 0 \cr \ast & \ast\end{array}\right)^2=\left(\begin{array}{cc} M_{2s-1}^2 & 0 \cr \star & \star\end{array}\right).$$
This implies $M_{2s-1}^2=0$, and so
$$\left(M_{2s-1}\Upsilon_{1}^{[0,2s-1)},M_{2s-1}\Upsilon_{2}^{[0,2s-1)},\ldots,M_{2s-1}\Upsilon_{2s-1}^{[0,2s-1)}\right)
=M_{2s-1}M_{2s-1}=0.$$
From this, we obtain $M_{2s-1}\Upsilon_{j}^{[0,2s-1)}=0$ for all $j$. Then by Eq. (\ref{eq3}), we have
\begin{equation}
\label{eq4}
\Upsilon_{j}^{[0,2s-1)}\in\mathcal{S}_{2s-1}, \ \forall j: \ 1\leq j\leq 2s-1.
\end{equation}

\par
  Now, from $M_{2s-1}M_{2s-1}=0$ and by linear algebra theory,
we deduce that
$$2\cdot {\rm rank}(M_{2s-1})={\rm rank}(M_{2s-1})+{\rm rank}(M_{2s-1})\leq 2s-1.$$
As $2\cdot {\rm rank}(M_{2s-1})$ is even, we have $2\cdot {\rm rank}(M_{2s-1})\leq 2s-2=2(s-1)$ and hence
${\rm rank}(M_{2s-1})\leq s-1$. On the other hand, by Eqs. (\ref{eq1}) and (\ref{eq2}) we see that
$M_{2s-1}$ is a strictly lower triangular  matrix and
\begin{eqnarray*}
M_{2s-1}
 &=&\left(\begin{array}{cccccc}
\left(\begin{array}{cc} 0 & 0 \cr 1 & 0\end{array}\right) & & & & \cr
\left(\begin{array}{cc} \ast & 0 \cr \ast & \ast\end{array}\right) & \left(\begin{array}{cc} 0 & 0 \cr 1 & 0\end{array}\right)& & &\cr
 \vdots &  \ddots & \ddots & & \cr
\star & \ldots &
\left(\begin{array}{cc} \ast & 0 \cr \ast & \ast\end{array}\right) &
\left(\begin{array}{cc} 0 & 0 \cr 1 & 0\end{array}\right) & \cr
\star &  \ldots & \star & (\ast,0) & 0
\end{array}\right) \\
 &=& \left(
\begin{array}{ccccccccc} 0 & & & & & & & & \cr
\textbf{1} & 0 & & & & & & &  \cr
\ast & 0 & 0 & & & & &  \cr
\ast & \ast & \textbf{1} & 0 & & & &  \cr
\vdots & \vdots & \vdots & \ddots  & \ddots & & & \cr
\ast & \ast &  \ast & \ldots  & 0 & 0 & &  &  \cr
\ast & \ast &  \ast & \ldots  & \ast & \textbf{1} & 0 &  &  \cr
\ast & \ast &  \ast & \ldots  & \ast & \ast & 0 &  & 0
\end{array}\right).
\end{eqnarray*}
This implies that $\left\{\Upsilon_{2i-1}^{[0,2s-1)}\mid 1\leq i\leq s-1\right\}$ is an
$\mathbb{F}_{2^m}$-linear independent subset of all column vectors
of the matrix $M_{2s-1}$, and hence
$${\rm rank}(M_{2s-1})\geq {\rm rank}\left\{\Upsilon_{2i-1}^{[0,2s-1)}\mid 1\leq i\leq s-1\right\}=s-1.$$
Summarizing the discussion above, we conclude that ${\rm rank}(M_{2s-1})=s-1$. Therefore,
$${\rm dim}_{\mathbb{F}_{2^m}}(\mathcal{S}_{2s-1})=(2s-1)-{\rm rank}(M_{2s-1})=s$$
and that $\left\{\Upsilon_{2i-1}^{[0,2s-1)}\mid 1\leq i\leq s-1\right\}$ is an
$\mathbb{F}_{2^m}$-linear independent subset of $\mathcal{S}_{2s-1}$ by Eq. (\ref{eq4}).

\par
  Moreover, by
  $M_{2s-1}\varepsilon_{2s-1}
  =M_{2s-1}(0,\ldots, 0,1)^{{\rm tr}}=0$, we have
$\varepsilon_{2s-1}\in \mathcal{S}_{2s-1}$.
It is clear that the vectors $\Upsilon_{1}^{[0,2s-1)}, \Upsilon_{3}^{[0,2s-1)},\ldots,\Upsilon_{2s-3}^{[2s-3,2s-1)}, \varepsilon_{2s-1}$
are linearly independent over $\mathbb{F}_{2^m}$. From this and by
${\rm dim}_{\mathbb{F}_{2^m}}(\mathcal{S}_{2s-1})=s$, we deduce that
$\{\Upsilon_{1}^{[0,2s-1)}, \Upsilon_{3}^{[0,2s-1)},\ldots,\Upsilon_{2s-3}^{[2s-3,2s-1)}, \varepsilon_{2s-1}\}$
is an $\mathbb{F}_{2^m}$-basis of $\mathcal{S}_{2s-1}$. Hence, we have proved the conclusion in (i).

\par
  Finally, the conclusion in (ii) follows from (i) immediately.
\hfill  $\Box$

\vskip 3mm
\par
  Now, we consider to construct all self-dual cyclic codes of length $2^k$ over ${\rm GR}(4,m)$. To do this, we use the following
notation:

\par
  Let $\mathbb{Z}_4=\{0,1,2,3\}$ in which the arithmetic is done modulo $4$, and let
$\mathbb{Z}_2=\{0,1\}$ in which the arithmetic is done modulo $2$. It is well known that $\mathbb{Z}_2$
is not a subfield of the ring $\mathbb{Z}_4$. In order to reduce the number of symbols,
we will regard $\mathbb{Z}_2$ as a subset of $\mathbb{Z}_4$. In that sense, we have that
$2\mathbb{Z}_2=\{0,2\}\subseteq \mathbb{Z}_4$ and each element $a$ in $\mathbb{Z}_4$ has a unique
$2$-adic expansion: $a=b_0+2b_1$, where $b_0,b_1\in \mathbb{Z}_2$. Define $\overline{a}=b_0=a$ (mod $2$). Then
$^{-}$ is a surjective homomorphism of rings from $\mathbb{Z}_4$ onto $\mathbb{Z}_2$. This homomorphism
can be naturally extended to a surjective homomorphism of polynomial rings from $\mathbb{Z}_4[z]$ onto $\mathbb{Z}_2[z]$
by:
$$\overline{a}(z)=\sum_{0\leq i\leq d}\overline{a}_iz^i\in \mathbb{Z}_2[z],
\ \forall a(z)=\sum_{0\leq i\leq d}a_iz^i\in \mathbb{Z}_4[z].$$
A monic polynomial $a(z)$ in $\mathbb{Z}_4[z]$ of positive degree is said to be \textit{basic irreducible}
if $\overline{a}(z)$ is an irreducible polynomial in $\mathbb{Z}_2[z]$.

\par
  From now on, let $m$ be a positive integer and let $\varsigma(z)$ be a fixed
monic basic irreducible polynomial in $\mathbb{Z}_4[z]$ of degree $m$. Then $\overline{\varsigma}(z)$
is an irreducible polynomial in $\mathbb{Z}_2[z]$ of degree $m$. We adopt the following notation:
\begin{description}
\item{$\bullet$}
  $R={\rm GR}(4,m)=\frac{\mathbb{Z}_4[z]}{\langle \varsigma(z)\rangle}
=\{\sum_{i=0}^{m-1}a_iz^i\mid a_0,a_1,\ldots,a_{m-1}\in \mathbb{Z}_4\}$
in which the arithmetic is done modulo $\varsigma(z)$.

\item{$\bullet$}
  $\mathbb{F}_{2^m}=\frac{\mathbb{Z}_2[z]}{\langle \overline{\varsigma}(z)\rangle}
=\{\sum_{i=0}^{m-1}b_iz^i\mid b_0,b_1,\ldots,b_{m-1}\in \mathbb{Z}_2\}=\overline{R}$
in which the arithmetic is done modulo $\overline{\varsigma}(z)$.
\end{description}

\noindent
Then $\mathbb{F}_{2^m}$ is a finite field of $2^m$ elements and $R$ is a Galois ring
of $4^m$ elements (cf. \cite[Theorem 14.1]{s10}).

\par
   As we have regarded $\mathbb{Z}_2$ as a subset of $\mathbb{Z}_4$,
we will regard $\mathbb{F}_{2^m}$ as a subset of $R$ in the natural way, though $\mathbb{F}_{2^m}$ is not a subfield
of $R$.

\par
  Let $\alpha=\sum_{i=0}^{m-1}a_iz^i\in R$, where
$a_i=b_{i0}+2b_{i1}\in \mathbb{Z}_4$ with $b_{i0}, b_{i1}\in \mathbb{Z}_2$ for all $i=0,1,\ldots,m-1$.
Then $\alpha$ has a unique $2$-adic expansion:
$$\alpha=\beta_0+2\beta_1, \ {\rm where} \ \beta_j=\sum_{i=0}^{m-1}b_{ij}z^i\in \mathbb{F}_{2^m}
\ {\rm for} \ j=0,1.$$
We define
$\overline{\alpha}=\beta_0=\sum_{i=0}^{m-1}\overline{a}_iz^i$ for all $\alpha \in R$.
Then the map $^{-}$ is a surjective homomorphism of rings from
$R$ onto $\mathbb{F}_{2^m}$ with kernel $2R=2\mathbb{F}_{2^m}=\{2\beta\mid \beta\in \mathbb{F}_{2^m}\}$. Here we only regard $\mathbb{F}_{2^m}$ as a subset of the ring $R$, but
$\mathbb{F}_{2^m}$ is not a subfield of $R$. In that sense, we get
$\mathbb{F}_{2^m}=\overline{R}=\{\overline{\alpha}\mid \alpha\in R\}$.

\par
  Let $k$ be any fixed positive integer. Then cyclic codes of length $2^k$ over the Galois ring $R$ are viewed
as ideals in the following ring:
\begin{description}
\item{$\bullet$}
 $\frac{R[x]}{\langle x^{2^k}-1\rangle}=R[x]/\langle x^{2^k}-1\rangle=\{\sum_{i=0}^{2^k-1}a_ix^i\mid
 a_0,a_1,\ldots,a_{2^k-1}\in R\}$ in which the arithmetic is done modulo $x^{2^k}-1$.
\end{description}
Namely, any vector
$(a_0,a_1,\ldots,a_{2^k-1})\in R^{2^k}$ is associated with the polynomial
$a_0+a_1x+\ldots+a_{2^k-1}x^{2^k-1}\in \frac{R[u]}{\langle x^{2^k}-1\rangle}$.

   Further, we set
\begin{description}
\item{$\bullet$}
 $\frac{\mathbb{F}_{2^m}[x]}{\langle x^{2^k}-1\rangle}=\mathbb{F}_{2^m}[x]/\langle x^{2^k}-1\rangle=\{\sum_{i=0}^{2^k-1}b_ix^i\mid
 b_0,b_1,\ldots,b_{2^k-1}\in \mathbb{F}_{2^m}\}$ in which the arithmetic is done modulo $x^{2^k}-1$.
\end{description}

\par
   As we have regarded $\mathbb{F}_{2^m}$ as a subset of $R={\rm GR}(4,m)$,
we will regard $\frac{\mathbb{F}_{2^m}[x]}{\langle x^{2^k}-1\rangle}$ as a subset of $\frac{R[x]}{\langle x^{2^k}-1\rangle}$ in the natural way, though $\frac{\mathbb{F}_{2^m}[x]}{\langle x^{2^k}-1\rangle}$ is not a subring
of $\frac{R[x]}{\langle x^{2^k}-1\rangle}$. In that sense, each element $\xi$ of $\frac{R[x]}{\langle x^{2^k}-1\rangle}$ has a unique
$2$-adic expansion:
$$\xi=\xi_0+2\xi_1, \ {\rm where} \ \xi_0, \xi_1\in \mathbb{F}_{2^m}[x]/\langle x^{2^k}-1\rangle.$$
This implies $2\cdot\frac{R[x]}{\langle x^{2^k}-1\rangle}=2\cdot\frac{\mathbb{F}_{2^m}[x]}{\langle x^{2^k}-1\rangle}
=\{2\xi_0\mid \xi_0\in \frac{\mathbb{F}_{2^m}[x]}{\langle x^{2^k}-1\rangle}\}$. Here we only regard $\frac{\mathbb{F}_{2^m}[x]}{\langle x^{2^k}-1\rangle}$ as a subset of
$\frac{R[x]}{\langle x^{2^k}-1\rangle}$ in the sense that $2\cdot 1=2\in \mathbb{Z}_4$ for $1\in \mathbb{Z}_2$.

\par
   Let $f(x)\in \frac{R[x]}{\langle x^{2^k}-1\rangle}$ and
$g(x)\in \frac{\mathbb{F}_{2^m}[x]}{\langle x^{2^k}-1\rangle}\subset \frac{R[x]}{\langle x^{2^k}-1\rangle}$. In this paper, let
$\langle f(x), 2g(x)\rangle$ be the ideal in the ring $\frac{R[x]}{\langle x^{2^k}-1\rangle}$ generated
by $f(x)$ and $2g(x)$. Then we have
\begin{eqnarray*}
 &&\langle f(x), 2g(x)\rangle\\
 &=& \left\{a(x)f(x)+2b(x)g(x)\mid a(x)\in \frac{R[x]}{\langle x^{2^k}-1\rangle},
 b(x)\in \frac{\mathbb{F}_{2^m}[x]}{\langle x^{2^k}-1\rangle}\right\}.
\end{eqnarray*}

\par
   Now, we give a direct approach to construct all distinct self-dual cyclic codes of length $2^k$ over $R={\rm GR}(4,m)$.

\vskip 3mm \noindent
  {\bf Theorem 2}
  \textit{For any positive integer $k$, we have the following conclusions}:

\vskip 2mm \par
  $\diamondsuit$ \textit{If $k=1$, $\langle 2\rangle$ is the only self-dual cyclic code
of length $2$ over ${\rm GR}(4,m)$}.

\vskip 2mm \par
  $\diamondsuit$ \textit{If $k=2$, all distinct $1+2^m$ self-dual cyclic codes of length $4$ over ${\rm GR}(4,m)$ are given by}:
$$\langle 2\rangle, \ \langle (u-1)^3+2b, 2(u-1)\rangle
\ {\mbox where} \ b\in \mathbb{F}_{2^m} \ {\mbox arbitrary}.$$

\par
  $\diamondsuit$ \textit{Let $k\geq 3$. For any integers $\delta$ and $l$, where $1\leq \delta<l\leq 2^k-1$ and $l$ is odd, let $\Upsilon_{j}^{[0,l)}$
be the $j$th column vector of the matrix $M_l$ (see Eq. (\ref{eq2})) and define its truncated vector
$\Upsilon_{j}^{[\delta,l)}$ by}
\begin{equation}
\label{eq5}
\Upsilon_{j}^{[\delta,l)}=\left(\begin{array}{c} g_{\delta,j}\cr g_{\delta+1,j}\cr
\vdots \cr g_{l-1,j}\end{array}\right)\in \mathbb{F}_2^{l-\delta},
\ {\mbox when} \ \Upsilon_{j}^{[0,l)}=\left(\begin{array}{c} g_{0,j}\cr \vdots \cr g_{\delta-1,j}\cr\hline g_{\delta,j}\cr g_{\delta+1,j}\cr
\vdots \cr g_{l-1,j}\end{array}\right)
\end{equation}
\textit{for all $j=1,2,\ldots,l$. Then all distinct self-dual cyclic codes of length $2^k$ over $R$
are given by the following three cases}:

\vskip 2mm\par
  {\bf I}. \textit{$1$ code: $\langle 2\rangle$}.

\vskip 2mm\par
  {\bf II}. \textit{$\sum_{\nu=0}^{2^{k-2}-1}(2^m)^{\nu+1}$ codes:
$$\left\langle (x-1)^{2^k-2\nu-1}+2(x-1)^{2^{k-1}-2\nu-2}+2b_{2\nu+1}(x),2(x-1)^{2\nu+1}\right\rangle,$$
where $0\leq \nu\leq 2^{k-2}-1$ and $b_{2\nu+1}(x)$ is given by the following two subcases}:

\par
  $\diamond$
  \textit{When $\nu=0$, we have $b_1(x)=b$, where $b\in \mathbb{F}_{2^m}$ arbitrary}.

\par
  $\diamond$ \textit{When $1\leq \nu\leq 2^{k-2}-1$, we have}
  $$b_{2\nu+1}(x)=\sum_{j=2\nu+1}^{4\nu}b_{j}(x-1)^{j-2\nu},$$
\textit{where
$\left(\begin{array}{c}b_{2\nu+1}\cr b_{2\nu+2}\cr \vdots \cr b_{4\nu}\end{array}\right)
=\sum_{i=\nu+1}^{2\nu}a_{2i-1}\Upsilon_{2i-1}^{[2\nu+1,4\nu+1)}
+a_{4\nu}\left(\begin{array}{c} 0\cr \vdots \cr 0 \cr 1\end{array}\right)$
and $a_{2i-1},a_{4\nu}\in \mathbb{F}_{2^m}$ arbitrary, for all $i=\nu+1,\nu+2,\ldots, 2\nu$}.

\par
  {\bf III}. \textit{$\sum_{\nu=1}^{2^{k-2}-1}(2^m)^{\nu+1}$ codes:
$$\left\langle (x-1)^{2^k-2\nu}+2(x-1)^{2^{k-1}-2\nu-1}+2b_{2\nu}(x),2(x-1)^{2\nu}\right\rangle,$$
where $1\leq \nu\leq 2^{k-2}-1$ and $b_{2\nu}(x)$ is given by}
$$b_{2\nu}(x)=\sum_{j=2\nu-1}^{4\nu-2}b_{j}(x-1)^{j-2\nu+1}$$
\textit{in which
$\left(\begin{array}{c}b_{2\nu-1}\cr b_{2\nu}\cr \vdots \cr b_{4\nu-2}\end{array}\right)
=\sum_{i=\nu}^{2\nu-1}a_{2i-1}\Upsilon_{2i-1}^{[2\nu-1,4\nu-1)}
+a_{4\nu-2}\left(\begin{array}{c} 0\cr \vdots \cr 0 \cr 1\end{array}\right)$
and $a_{2i-1},a_{4\nu-2}$ $\in \mathbb{F}_{2^m}$ arbitrary, for all $i=\nu,\nu+1,\ldots, 2\nu-1$}.

\vskip 2mm\par
  \textit{Therefore, the number $N_{{\rm E}}({\rm GR}(4,m),2^k)$ of all
self-dual cyclic codes of length $2^k$ over ${\rm GR}(4,m)$ is}
$$N_{{\rm E}}({\rm GR}(4,m),2^k)=1+2^m+2(2^m)^2\left(\frac{(2^m)^{2^{k-2}-1}-1}{2^m-1}\right).$$

\noindent
 {\bf Remark}
  The above formula for the number of all
self-dual cyclic codes of length $2^k$ over ${\rm GR}(4,m)$ has been given in \cite[Corollary 3.5]{s7}.

\vskip 3mm\par
  Finally, we give an explicit expression for every
self-dual cyclic code of length $2^k$ over ${\rm GR}(4,m)$. From \cite[Proposition 2]{DM20}, we obtain
$$G_{2^k}=\left(\begin{array}{cccc} g_{1,1}^{(2^k)} & g_{1,2}^{(2^k)} & \ldots & g_{1,2^k}^{(2^k)} \cr
g_{2,1}^{(2^k)} & g_{2,2}^{(2^k)} & \ldots & g_{2,2^k}^{(2^k)}\cr
\ldots & \ldots & \ldots & \ldots \cr
g_{2^k,1}^{(2^k)} & g_{2^k,2}^{(2^k)} & \ldots & g_{2^k,2^k}^{(2^k)}\end{array}\right)
\ ({\rm mod} \ 2),$$
where
\begin{equation}
\label{eq6}
g_{i,j}^{(2^k)}=\left(\begin{array}{c}2^k-j\cr i-j\end{array}\right), \
g_{i,i}^{(2^k)}=1, \
\ \mbox{and} \ \left(\begin{array}{c}2^k-j\cr i-j\end{array}\right)=0
\ \mbox{if} \ i<j.
\end{equation}
Now, by Eqs. (\ref{eq1}), (\ref{eq2}) and (\ref{eq6}), we have the following theorem.
Using these explicit expressions of the codes in the theorem, one can further study the parametric properties of this class of self-dual cyclic codes.

\vskip 3mm\noindent
  {\bf Theorem 3} \textit{For any integer $k\geq 3$, all distinct self-dual cyclic codes of length $2^k$ over ${\rm GR}(4,m)$
are given by the following three cases}:

\vskip 2mm\par
  (i) \textit{$1$ code: $\langle 2\rangle$}.

\vskip 2mm\par
  (ii) \textit{$2^m$ codes: $\langle (x-1)^{2^k-1}+2b,2(x-1)\rangle$, where $b\in \mathbb{F}_{2^m}$
arbitrary}.

\vskip 2mm\par
  (iii) \textit{For each even integer $s$: $2\leq s\leq 2^{k-1}-1$, there are $2^{m(\lfloor\frac{s}{2}\rfloor+1)}$ codes:
$$\langle (x-1)^{2^k-s}+2(x-1)^{2^{k-1}-s-1}+2b_s(x),2(x-1)^s\rangle,$$
where
\begin{eqnarray*}
b_s(x)&=&\sum_{i=\lfloor\frac{s+1}{2}\rfloor}^{s-1}\sum_{t=1}^{2(s-i)}a_{2i-1}
   \left(\begin{array}{c}2^k-2i+1 \cr t\end{array}\right)(x-1)^{2i-1-s+t} \\
 && +a_{2s-2}(x-1)^{s-1} \ \ ({\rm mod} \ 2)
\end{eqnarray*}
and $a_{2i-1},a_{2s-2}\in \mathbb{F}_{2^m}$ arbitrary, for
all integers $i$}:
$\lfloor\frac{s+1}{2}\rfloor\leq i\leq s-1.$

\vskip 3mm
\noindent
  {\bf Proof}. Obviously, we only need to prove the conclusion in Case (iii).
For any integer $s$, $2\leq s\leq 2^{k-1}-1$, let
$\lfloor\frac{s+1}{2}\rfloor\leq i\leq s-1$. By the definition for the truncated vector $\Upsilon_{2i-1}^{[s-1,2s-1)}$
(see Eq. (\ref{eq5})) and Eq. (\ref{eq6}), we have that: \\
$\Upsilon_{2i-1}^{[s-1,2s-1)}=\left(\begin{array}{c}g_{s-1,2i-1} \cr g_{s,2i-1}\cr \vdots \cr g_{2s-2,2i-1}\end{array}\right)$
and $g_{s-1+\gamma,2i-1}$ satsifies the following conditions:
\begin{description}
\item{$\triangleright$}
   If $0\leq\gamma<2i-1-s$,
$g_{s-1+\gamma,2i-1}=\left(\begin{array}{c}2^k-(2i-1)\cr s+\gamma-(2i-1)\end{array}\right)=0$.

\item{$\triangleright$}
   If $\gamma=2i-1-s$,
$g_{s-1+\gamma,2i-1}=\left(\begin{array}{c}2^k-(2i-1)\cr s+\gamma-(2i-1)\end{array}\right)+1=0$.

\item{$\triangleright$}
  If $\gamma=2i-1-s+t$, where $1\leq t\leq 2(s-i)$,
  $$g_{s-1+\gamma,2i-1}=\left(\begin{array}{c}2^k-(2i-1)\cr t\end{array}\right)
  =\left(\begin{array}{c}2^k-2i+1\cr t\end{array}\right).$$
\end{description}
From these and by Theorem 2, we deduce the conclutions in Case (iii) directly.
Here, we omit the trivial verification process.
\hfill $\Box$



\section{Proof of Theorem 2}
\noindent
 In this section, we give a proof for Theorem 2.
First, from \cite[Lemma 3.1]{s6}, we deduce the following lemma.

\vskip 3mm
\noindent
  {\bf Lemma 1}
  \textit{In the ring $\frac{R[x]}{\langle x^{2^k}-1\rangle}$, we have $(x-1)^{2^k}=2(x-1)^{2^{k-1}}$}.

\vskip 3mm
\noindent
  {\bf Lemma 2}
  \textit{Let $1\leq s\leq 2^{k-1}$. For any vector
$\underline{b}=(b_0,b_1,\ldots,b_{s-1})^{{\rm tr}}\in\mathbb{F}_{2^m}^s$, we set
$b(x)=\sum_{j=0}^{s-1}b_j(x-1)^j$ and let $\mathcal{C}_{\underline{b}}$
be an ideal of $\frac{R[x]}{\langle x^{2^k}-1\rangle}$ defined by
$$\mathcal{C}_{\underline{b}}=\langle (x-1)^{2^k-s}+2b(x),2(x-1)^s\rangle.$$
Then we have the following}:
\begin{itemize}
 \item[(i)]
  (cf. \cite[Theorems 2.2 and 3.8]{s6}) \textit{The ideal $\mathcal{C}_{\underline{b}}$ is a cyclic code of length $2^k$ over $R$ containing
$(2^m)^{2^k}$ codewords}.

\item[(ii)]
 \textit{We have
$\mathcal{C}_{\underline{b}}\neq \mathcal{C}_{\underline{c}}$ for any $\underline{b}, \underline{c}\in \mathbb{F}_{2^m}^s$
satisfying $\underline{b}\neq \underline{c}$}.
\end{itemize}

\vskip 3mm
\noindent
   {\bf Proof.}
    (ii) Let $\underline{b}=(b_0,b_1,\ldots,b_{s-1})^{{\rm tr}}, \underline{c}=(c_0,c_1,\ldots,c_{s-1})^{{\rm tr}}\in\mathbb{F}_{2^m}^s$
and $\underline{b}\neq \underline{c}$.
We set $b(x)=\sum_{j=0}^{s-1}b_j(x-1)^j$ and $c(x)=\sum_{j=0}^{s-1}c_j(x-1)^j$. Suppose that
$\mathcal{C}_{\underline{b}}=\mathcal{C}_{\underline{c}}$. Then
$$2(b(x)-c(x))=\left((x-1)^{2^k-s}+2b(x)\right)-\left((x-1)^{2^k-s}+2c(x)\right)
\in \mathcal{C}_{\underline{b}}.$$
This implies $b(x)-c(x)\in \langle (x-1)^s\rangle$, i.e., $b(x)\equiv c(x)$ (mod $(x-1)^s$)
in the polynomial ring $\mathbb{F}_{2^m}[x]$. From this, we deduce that $\underline{b}=\underline{c}$, which is a contradiction.
That proves $\mathcal{C}_{\underline{b}}\neq \mathcal{C}_{\underline{c}}$.
\hfill  $\Box$

\vskip 3mm
\par
  As $x^{2^k}=1$ in the rings $\frac{R[x]}{\langle x^{2^k}-1\rangle}$ and $\frac{\mathbb{F}_{2^m}[x]}{\langle x^{2^k}-1\rangle}$,
we have $x^{-l}=x^{2^k-l}$ for all integers $l$, $1\leq l\leq 2^k-1$.

\vskip 3mm
\noindent
  {\bf Lemma 3}
 \textit{Let $1\leq s\leq 2^{k-1}$. For any vector
$\underline{b}=(b_0,b_1,\ldots,b_{s-1})^{{\rm tr}}\in\mathbb{F}_{2^m}^s$,
let $b(x)=\sum_{j=0}^{s-1}b_j(x-1)^j$ and set
$\mathcal{C}_{\underline{b}}=\langle (x-1)^{2^k-s}+2b(x),2(x-1)^s\rangle\subseteq \frac{R[x]}{\langle x^{2^k}-1\rangle}$.
Then $\mathcal{C}_{\underline{b}}$ is a self-dual cyclic
code of length $2^k$ over $R$, if $b(x)$ satisfies the following
congruence relation over the ring $\mathbb{F}_{2^m}[x]$}:
\begin{equation}
\label{eq7}
 b(x)+x^{-s}b(x^{-1})\equiv (x-1)^{2^{k-1}-s}
\ ({\rm mod} \ (x-1)^s).
\end{equation}

\vskip 3mm
\noindent
   {\bf Proof.}
    Let $b(x)$ satisfy Eq. (\ref{eq7}).
By Lemma 2 (i), we know that
$\mathcal{C}_{\underline{b}}$ is a cyclic
code of length $2^k$ over $R$ containing $(2^m)^{2^k}=(|R|^{2^k})^{\frac{1}{2}}$
codewords. Recall that the
annihilator of $\mathcal{C}_{\underline{b}}$ is defined by
$${\rm Ann}(\mathcal{C}_{\underline{b}})=\left\{a(x)\in R[x]/\langle x^{2^k}-1\rangle\mid a(x)c(x)=0, \ \forall c(x)\in \mathcal{C}_{\underline{b}}\right\}.$$
Let $\chi: \frac{R[x]}{\langle x^{2^k}-1\rangle}\rightarrow \frac{R[x]}{\langle x^{2^k}-1\rangle}$ be a conjugate map
defined by
$$\chi(a(x))=a(x^{-1})=a_0+\sum_{i=1}^{2^k-1}a_ix^{2^k-i}, \
\forall a(x)=\sum_{i=0}^{2^k-1}a_ix^{i} \ {\rm where}
\ a_i\in R.$$
Then it is well known that (cf. \cite[Theorem 4.1]{s6}) the dual code $\mathcal{C}_{\underline{b}}^{\bot}$ of $\mathcal{C}_{\underline{b}}$
is given by
$$\mathcal{C}_{\underline{b}}^{\bot}=\chi({\rm Ann}(\mathcal{C}_{\underline{b}}))=\{\chi(a(x))\mid a(x)\in {\rm Ann}(\mathcal{C}_{\underline{b}})\}.$$

\par
  Now, by Eq. (\ref{eq7}) there exists $g(x)\in \frac{\mathbb{F}_{2^m}[x]}{\langle x^{2^k}-1\rangle}$ such that
$$x^{-s}b(x^{-1})=b(x)+(x-1)^{2^{k-1}-s}+g(x)(x-1)^s.$$ This implies
$$2x^{-s}b(x^{-1})=2\left(b(x)+(x-1)^{2^{k-1}-s}\right)+g(x)\cdot 2(x-1)^s
\ {\rm in} \ R[x]/\langle x^{2^k}-1\rangle.$$
Since $x$ is invertible in $\frac{R[x]}{\langle x^{2^k}-1\rangle}$, by $-2=2$ in $\mathbb{Z}_4\subset R$, it follows that
\begin{eqnarray*}
\chi(\mathcal{C}_{\beta}) &=& \langle \chi((x-1)^{2^k-s}+2b(x)),\chi(2(x-1)^s)\rangle \\
 &=& \langle (x^{-1}-1)^{2^k-s}+2b(x^{-1}), 2(x^{-1}-1)^s\rangle\\
 &=& \langle (-1)^{2^k-s}x^{-(2^k-s)}(x-1)^{2^k-s}+2b(x^{-1}), 2x^{-s}(u-1)^s\rangle\\
 &=& \langle (x-1)^{2^k-s}+2x^{-s}b(x^{-1}),2(x-1)^s\rangle\\
 &=& \langle (x-1)^{2^k-s}+2(b(x)+(x-1)^{2^{k-1}-s})+g(x)\cdot 2(x-1)^s, \\
   && 2(x-1)^s\rangle \\
 &=& \langle (x-1)^{2^k-s}+2(b(x)+(x-1)^{2^{k-1}-s}),2(x-1)^s\rangle.
\end{eqnarray*}

\par
  Moreover, by Lemma 1 we have that $(x-1)^{2^k-s}\cdot 2(x-1)^s=2(x-1)^{2^k}=2\cdot 2(x-1)^{2^{k-1}}=0$.
Similarly, by $2^k-2s\geq 2^k-2\cdot 2^{k-1}\geq 0$, it follows that
$$(x-1)^{2(2^k-s)}=(x-1)^{2^k}(x-1)^{2^k-2s}=2(x-1)^{2^{k-1}}(x-1)^{2^k-2s},$$
and hence
\begin{eqnarray*}
 &&\left((x-1)^{2^k-s}+2b(x)\right)\left((x-1)^{2^k-s}+2\left(b(x)+(x-1)^{2^{k-1}-s}\right)\right)\\
 &=&(x-1)^{2(2^k-s)}+2(x-1)^{2^k-s}\left(b(x)+b(x)+(x-1)^{2^{k-1}-s}\right)\\
 &=&2(x-1)^{2^{k-1}+2^k-2s}+2(x-1)^{2^k-s}\cdot(x-1)^{2^{k-1}-s}\\
 &=&0.
\end{eqnarray*}
From these, we deduce that $\mathcal{C}_{\underline{b}}\cdot \chi(\mathcal{C}_{\underline{b}})=\{0\}$. Since
$R$ is a Galois ring and $|\chi(\mathcal{C}_{\underline{b}})|=|\mathcal{C}_{\underline{b}}|=(|R|^{2^k})^{\frac{1}{2}}$, we conclude that
${\rm Ann}(\mathcal{C}_{\underline{b}})=\chi(\mathcal{C}_{\underline{b}})$.

\par
  Finally, by $\chi^{-1}=\chi$, we have $\mathcal{C}_{\underline{b}}^{\bot}=\chi({\rm Ann}(\mathcal{C}_{\underline{b}}))
  =\chi^2(\mathcal{C}_{\underline{b}})=\mathcal{C}_{\underline{b}}$.
\hfill $\Box$

\vskip 3mm
\par
  From \cite[Theorem 1 (i) and (ii)]{s3} and by $-1=1$ in
the finite field $\mathbb{F}_{2^m}$, we deduce the following lemma.

\vskip 3mm
\noindent
  {\bf Lemma 4}
  \textit{Let $l$ be an integer satisfying $1\leq l\leq 2^k-1$. Let
$B_l=(b_0,b_1,\ldots$, $b_{l-1})^{{\rm tr}}\in \mathbb{F}_{2^m}^l$
and set $\beta(x)=\sum_{j=0}^{l-1}b_j(x-1)^j$. Using the notation of Theorem 1, the polynomial $\beta(x)$
satisfies the following congruence relation}:
$$
\beta(x)+x^{-1}\beta(x^{-1})\equiv 0 \ ({\rm mod} \ (x-1)^l)
$$
\textit{in the ring $\mathbb{F}_{2^m}[x]$, where $x^{-1}=x^{2^k-1}$ (mod $(x-1)^l$), if and only if $B_l$ is a solution vector of the system of linear equations}
$$M_{l}(y_0, y_1, \ldots, y_{l-1})^{{\rm tr}}=(0,0, \ldots,0)^{{\rm tr}}.$$

\par
  The following result plays a key role in this paper.

\vskip 3mm
\noindent
  {\bf Lemma 5}
 \textit{For any integer $s$: $2^{k-2}+1\leq s\leq 2^{k-1}-1$ where $k\geq 3$, we set
$$\rho_k(x)=(x-1)^{2^{k-1}-s-1}.$$
 Then $\rho_k(x)$ satisfies Eq. (\ref{eq7}) in Lemma 3, i.e.,}
$$
 \rho_k(x)+x^{-s}\rho_k(x^{-1})\equiv (x-1)^{2^{k-1}-s}
\ ({\rm mod} \ (x-1)^s).
$$

\noindent
  {\bf Proof.}
  Let $\varphi(x)=(x-1)^{s-1}\rho_k(x)=(x-1)^{2^{k-1}-2}$. As the characteristic of the finite field $\mathbb{F}_{2^m}$ is $2$, we have
\begin{eqnarray*}
x^{2^{k-1}+1}&=&\left(1+(x-1)\right)^{2^{k-1}}(1+(x-1))=\left(1+(x-1)^{2^{k-1}}\right)(1+(x-1))\\
 &=&1+(x-1)+(x-1)^{2^{k-1}}x.
\end{eqnarray*}
As $s\leq 2^{k-1}-1$, we have $2s-1\leq 2^{k}-3$. This implies
$$(x-1)^{2^{k-1}-2}(x-1)^{2^{k-1}}x=(x-1)^{2^k-2}x\equiv 0 \ ({\rm mod} \ (x-1)^{2s-1}).$$
From these and by $(x^{-1})^{2^{k-1}-2}=x^{2^k-(2^{k-1}-2)}=x^{2^{k-1}+2}$ (mod $(x-1)^{2^k}$), we deduce
\begin{eqnarray*}
\varphi(x)+x^{-1}\varphi(x^{-1})
 &=&(x-1)^{2^{k-1}-2}+x^{-1}(x^{-1}-1)^{2^{k-1}-2}\\
 &=&(x-1)^{2^{k-1}-2}+x^{2^{k-1}+1}(x-1)^{2^{k-1}-2}\\
 &=&(x-1)^{2^{k-1}-2}\left(1+x^{2^{k-1}+1}\right) \\
 &\equiv & (x-1)^{2^{k-1}-2}(1+(1+(x-1)))\\
 &\equiv & (x-1)^{2^{k-1}-1}  \ ({\rm mod} \ (x-1)^{2s-1}).
\end{eqnarray*}
Then by $\varphi(x)=(x-1)^{s-1}\rho_k(x)$, $(x-1)^{2s-1}=(x-1)^{s-1}\cdot (x-1)^s$ and
$$x^{-1}\varphi(x^{-1})=x^{-1}(x^{-1}-1)^{s-1}\rho_k(x^{-1})=(x-1)^{s-1}\cdot x^{-s}\rho_k(x^{-1}),$$
we conclude that $\rho_k(x)$ satisfies Eq. (\ref{eq7}) in Lemma 3.
\hfill $\Box$

\vskip 3mm
\par
  Now, we prove Theorem 2 in Section 2. It is obvious that $\langle 2\rangle=2\cdot \frac{R[x]}{\langle x^{2^k}-1\rangle}$ is a trivial self-dual cyclic code of length $2^k$ over the Galois ring $R={\rm GR}(4,m)$.

\begin{center}
 Case $k=1$
\end{center}

\par
  As $1\leq s\leq 2^{k-1}=1$, we have $s=1$. Since there is no element $b(u)=b\in\mathbb{F}_{2^m}$
satisfying Eq. (\ref{eq7}) in Lemma 3, $\langle 2\rangle$ is the only
self-dual cyclic code of length $2$ over ${\rm GR}(4,m)$.

\begin{center}
 Case $k=2$
\end{center}

\par
  \par
   As $1\leq s\leq 2^{k-1}=2$, we have $s=1,2$. When $s=2$, we have $x^2\equiv1$ (mod $(x-1)^2$). This implies
$x^{-1}\equiv x$ (mod $(x-1)^2$). In this case, we have
$$b(x)+x^{-2}b(x^{-1})\equiv 0 \not\equiv (x-1)^{2^{2-1}-2}
\ ({\rm mod} \ (x-1)^2),$$
for any $b(x)=b_0+b_1(x-1)$ with $b_0,b_1\in \mathbb{F}_{2^m}$. Hence there is no polynomial
$b(x)\in \frac{\mathbb{F}_{2^m}[x]}{\langle (x-1)^2\rangle}$ satisfying
Eq. (\ref{eq7}), when $s=2$.

\par
  Let $s=1$. Then $x\equiv1$ (mod $(x-1)^1$). For any $b(x)=b\in\mathbb{F}_{2^m}$, we have
$$b(x)+x^{-1}b(x^{-1})\equiv x+x \equiv 0 \equiv (x-1)^{2^{2-1}-1}  \ ({\rm mod} \ x-1).$$

\par
  Therefore, there are $1+2^m$ self-dual cyclic codes of length $2^2$ over ${\rm GR}(4,m)$:
$$\langle 2\rangle, \ \langle (x-1)^3+2b, 2(x-1)\rangle
\ {\rm where} \ b\in \mathbb{F}_{2^m} \ {\rm arbitrary}.$$

\begin{center}
 Case $k\geq 3$
\end{center}

\par
  In this situation, we consider two cases for nontrivial self-dual cyclic codes of length $2^k$ over $R$: when $s$ is odd and when $s$ is even.

\par
  Case II: Let $s=2\nu+1$,
where $0\leq \nu\leq 2^{k-2}-1$. We further split this case into three subcases.

\par
  \textsl{Subcase} 1:  Let $\nu=0$. Then $s=1$. In this case, by Lemma 3,
the code $\langle (x-1)^{2^k-1}+2b,2(x-1)\rangle$ is a self-dual cyclic code of length $2^k$
over ${\rm GR}(4,m)$, for any $b\in \mathbb{F}_{2^m}$. All these
$2^m$ codes are distinct by Lemma 2 (ii).

\par
   \textsl{Subcase} 2:  Let $1\leq \nu\leq 2^{k-2}-1$. By Eqs. (\ref{eq1}) and (\ref{eq2}),
we see that $M_{4\nu+1}$ is a strictly lower triangular matrix. By Eq. (\ref{eq5}), the $(2i-1)$st
column vector of the matrix $M_{4\nu+1}$ is equal to
\begin{equation}
\label{eq8}
\Upsilon_{2i-1}^{[0,4\nu+1)}=\left(\begin{array}{c}\textbf{0}_{(2\nu+1)\times 1}\cr\hline \Upsilon_{2i-1}^{[2\nu+1,4\nu+1)}\end{array}\right),
\ \forall i: \ \nu+1\leq i\leq 2\nu,
\end{equation}
where $\textbf{0}_{(2\nu+1)\times 1}$ is the zero matrix of size $(2\nu+1)\times 1$. Denote by
$\mathcal{S}_{4\nu+1}^{(2\nu+1)}$ the subset of column vectors in $\mathbb{F}_{2^m}^{4\nu+1}$ defined by:
$$\sum_{\nu+1\leq i\leq 2\nu}a_{2i-1}\Upsilon_{2i-1}^{[0,4\nu+1)}+a_{4\nu}\varepsilon_{4\nu+1},
\ {\rm where} \ a_{2i-1}, a_{4\nu}\in \mathbb{F}_{2^m}, \ \nu+1\leq i\leq 2\nu.$$
By Theorem 1, we see that $\{\Upsilon_{2\nu+1}^{[0,4\nu+1)}, \Upsilon_{2\nu+3}^{[0,4\nu+1)},\ldots,
\Upsilon_{4\nu-1}^{[0,4\nu+1)}, \varepsilon_{4\nu+1}\}$ is an $\mathbb{F}_{2^m}$-linearly independent subset
of $\mathcal{S}_{4\nu+1}$. This implies that
$$\mathcal{S}_{4\nu+1}^{(2\nu+1)}\subseteq \mathcal{S}_{4\nu+1} \ {\rm and}
\ |\mathcal{S}_{4\nu+1}^{(2\nu+1)}|=(2^m)^{\nu+1}.$$

\par
  Now, let $B_{4\nu+1}=(b_0,b_1,\ldots,b_{4\nu})^{{\rm tr}}\in \mathcal{S}_{4\nu+1}^{(2\nu+1)}$. Then
there exists a unique vector $(a_{2\nu+1},a_{2\nu+3},\ldots, a_{4\nu-1},a_{4\nu})^{{\rm tr}}\in \mathbb{F}_{2^m}^{\nu+1}$
such that
$$B_{4\nu+1}=\sum_{\nu+1\leq i\leq 2\nu}a_{2i-1}\Upsilon_{2i-1}^{[0,4\nu+1)}+a_{4\nu}\varepsilon_{4\nu+1}.$$
From this and by Eq. (\ref{eq8}), we deduce that $b_j=0$ for all integers $j$: $0\leq j\leq 2\nu$, and
$$\left(\begin{array}{c}b_{2\nu+1}\cr b_{2\nu+2}\cr \vdots \cr b_{4\nu}\end{array}\right)
=\sum_{i=\nu+1}^{2\nu}a_{2i-1}\Upsilon_{2i-1}^{[2\nu+1,4\nu+1)}
+a_{4\nu}\left(\begin{array}{c} 0\cr \vdots \cr 0 \cr 1\end{array}\right).$$
Then we define
$$b_{2\nu+1}(x)=\sum_{j=2\nu+1}^{4\nu}b_j(x-1)^{j-2\nu}
\ {\rm and} \ \beta(x)=\sum_{j=0}^{4\nu}b_j(x-1)^j=(x-1)^{2\nu}b_{2\nu+1}(x).$$

\par
  As $B_{4\nu+1}\in \mathcal{S}_{4\nu+1}$, by Lemma 4 it follows that
$$\beta(x)+x^{-1}\beta(x^{-1})\equiv 0 \ ({\rm mod} \ (x-1)^{4\nu+1}).$$
Then by $\beta(x^{-1})=(x^{-1}-1)^{2\nu}b_{2\nu+1}(x^{-1})=(x-1)^{2\nu}\cdot x^{-2\nu}b_{2\nu+1}(x^{-1})$, we have
$$
b_{2\nu+1}(x)+x^{-(2\nu+1)}b_{2\nu+1}(x^{-1})\equiv 0 \ ({\rm mod} \ (x-1)^{2\nu+1}).
$$
From this and by Lemma 5, we deduce that
\begin{eqnarray*}
&&\left(\rho_k(x)+b_{2\nu+1}(x)\right)+x^{-(2\nu+1)}\left(\rho_k(x^{-1})+b_{2\nu+1}(x^{-1})\right)\\
&\equiv & (x-1)^{2^{k-1}-(2\nu+1)} \ ({\rm mod} \ (x-1)^{2\nu+1}),
\end{eqnarray*}
where $\rho_k(x)=(x-1)^{(2^{k-1}-1)-(2\nu+1)}=(x-1)^{2^{k-1}-2\nu-2}$. Therefore, using Lemma 3, we conclude that
the following ideal in $\frac{R[x]}{\langle x^{2^k}-1\rangle}$:
$$\mathcal{C}=\langle (x-1)^{2^k-2\nu-1}+2((x-1)^{2^{k-1}-2\nu-2}+b(x)),2(x-1)^{2\nu+1}\rangle$$
is a self-dual cyclic code of length $2^k$ over $R$.

\par
  As stated above, by Theorem 1 (ii) and $|\mathcal{S}_{4\nu+1}^{(2\nu+1)}|=(2^m)^{\nu+1}$, we conclude that
there are $\sum_{\nu=0}^{2^{k-2}-1}(2^m)^{\nu+1}$ distinct
self-dual cyclic codes of length $2^k$ over $R$ given by Case II of Theorem 2.

\par
  Case III: Let $s=2\nu$,
where $1\leq \nu\leq 2^{k-2}-1$. Since $M_{4\nu-1}$ is a strictly lower triangular matrix, by Eq. (\ref{eq5}) the $(2i-1)$st
column vector of the matrix $M_{4\nu-1}$ is equal to
\begin{equation}
\label{eq9}
\Upsilon_{2i-1}^{[0,4\nu-1)}=\left(\begin{array}{c}\textbf{0}_{(2\nu-1)\times 1}\cr\hline \Upsilon_{2i-1}^{[2\nu-1,4\nu-1)}\end{array}\right),
\ \forall i: \ \nu\leq i\leq 2\nu-1,
\end{equation}
where $\textbf{0}_{(2\nu-1)\times 1}$ is the zero matrix of size $(2\nu-1)\times 1$. Let
$\mathcal{S}_{4\nu-1}^{(2\nu-1)}$ be the subset of vectors in $\mathbb{F}_{2^m}^{4\nu-1}$ defined by:
$$\sum_{\nu\leq i\leq 2\nu-1}a_{2i-1}\Upsilon_{2i-1}^{[0,4\nu-1)}+a_{4\nu-2}\varepsilon_{4\nu-1},$$
where $a_{2i-1}, a_{4\nu-2}\in \mathbb{F}_{2^m}$ for all integers $i$: $\nu\leq i\leq 2\nu-1.$
By Theorem 1, we see that $\{\Upsilon_{2\nu-1}^{[0,4\nu-1)}, \Upsilon_{2\nu+1}^{[0,4\nu-1)},\ldots,
\Upsilon_{4\nu-3}^{[0,4\nu-1)}, \varepsilon_{4\nu-1}\}$ is an $\mathbb{F}_{2^m}$-linearly independent subset
of $\mathcal{S}_{4\nu-1}$. This implies that
$$\mathcal{S}_{4\nu-1}^{(2\nu-1)}\subseteq \mathcal{S}_{4\nu-1} \ {\rm and}
\ |\mathcal{S}_{4\nu-1}^{(2\nu-1)}|=(2^m)^{\nu+1}.$$

\par
  Now, let $B_{4\nu-1}=(b_0,b_1,\ldots,b_{4\nu-2})^{{\rm tr}}\in \mathcal{S}_{4\nu-1}^{(2\nu-1)}$. Then
there exists a unique vector $(a_{2\nu-1},a_{2\nu+1},\ldots, a_{4\nu-3},a_{4\nu-2})^{{\rm tr}}\in \mathbb{F}_{2^m}^{\nu+1}$
such that
$$B_{4\nu-1}=\sum_{\nu\leq i\leq 2\nu-1}a_{2i-1}\Upsilon_{2i-1}^{[0,4\nu-1)}+a_{4\nu-2}\varepsilon_{4\nu-1}.$$
From this and by Eq. (\ref{eq9}), we deduce that $b_j=0$ for all integers $j$: $0\leq j\leq 2\nu-2$, and
$$\left(\begin{array}{c}b_{2\nu-1}\cr b_{2\nu}\cr \vdots \cr b_{4\nu-2}\end{array}\right)
=\sum_{i=\nu}^{2\nu-1}a_{2i-1}\Upsilon_{2i-1}^{[2\nu-1,4\nu-1)}
+a_{4\nu-2}\left(\begin{array}{c} 0\cr \vdots \cr 0 \cr 1\end{array}\right).$$
Then we define $b_{2\nu}(x)=\sum_{j=2\nu-1}^{4\nu-2}b_j(x-1)^{j-2\nu+1}$ and set
$$\beta(x)=\sum_{j=0}^{4\nu-2}b_j(x-1)^j=(x-1)^{2\nu-1}b_{2\nu}(x).$$

\par
  As $B_{4\nu-1}\in \mathcal{S}_{4\nu-1}$, by Lemma 4 it follows that
$$\beta(x)+x^{-1}\beta(x^{-1})\equiv 0 \ ({\rm mod} \ (x-1)^{4\nu-1}).$$
From this and by
$$\beta(x^{-1})=(x^{-1}-1)^{2\nu-1}b(x^{-1})=(x-1)^{2\nu-1}\cdot x^{-(2\nu-1)}b_{2\nu}(x^{-1}),$$
we deduce that
$$
b(x)+x^{-2\nu}b(x^{-1})\equiv 0 \ ({\rm mod} \ (x-1)^{2\nu}).
$$
Let $\rho_k(x)=(x-1)^{2^{k-1}-2\nu-1}$. Then by Lemma 5, we have
\begin{eqnarray*}
 &&\left(\rho_k(x)+b_{2\nu}(x)\right)+x^{-2\nu}\left(\rho_k(x^{-1})+b_{2\nu}(x^{-1})\right) \\
 &\equiv & (x-1)^{2^{k-1}-2\nu} \ ({\rm mod} \ (x-1)^{2\nu}).
\end{eqnarray*}
Hence by Lemma 3, we see that the ideal in $\frac{R[x]}{\langle x^{2^k}-1\rangle}$:
$$\mathcal{C}=\langle (x-1)^{2^k-2\nu}+2((x-1)^{2^{k-1}-2\nu-1}+b_{2\nu}u)),2(x-1)^{2\nu}\rangle$$
is a self-dual cyclic code of length $2^k$ over $R$.

\par
  As stated above, by Theorem 1 (ii) and $|\mathcal{S}_{4\nu-1}^{(2\nu-1)}|=(2^m)^{\nu+1}$, we conclude that
there are $\sum_{\nu=1}^{2^{k-2}-1}(2^m)^{\nu+1}$ distinct
self-dual cyclic codes of length $2^k$ over $R$ given by Case III of Theorem 2.

\par
  Since $2\nu+1$ is odd and $2\nu$ is even, we see that there is no overlap between Cases II and III. Hence, the number of self-dual cyclic codes constructed above is equal to
$$1+\sum_{\nu=0}^{2^{k-2}-1}(2^m)^{\nu+1}+\sum_{\nu=1}^{2^{k-2}-1}(2^m)^{\nu+1}=N_{{\rm E}}({\rm GR}(4,m),2^k).$$

\par
  Therefore, all distinct self-dual cyclic codes of length $2^k$ over $R$
are given by Theorem 2. This proves the theorem.




\section{Self-dual cyclic codes over ${\rm GR}(4,m)$ of lengths $2^4$, $2^5$ and $2^6$}
\noindent
  In this section, we make a list of all distinct
  self-dual cyclic codes of length $2^k$ over the Galois ring $R={\rm GR}(4,m)$,
using Theorem 2 or Theorem 3. To save space, we only consider the cases $k=4,5,6$.

\vskip 3mm \noindent
 {\bf Example 1}
 All $1+2^m+2(2^{2m}+2^{3m}+2^{4m})$
self-dual cyclic codes of length $16$ over $R$ are given by the following two cases:

\par
   (i) $1$ code: $\langle 2\rangle$.

\par
   (ii) $2^m+2(2^{2m}+2^{3m}+2^{4m})$ codes:
$$\langle (x-1)^{16-s}+2(x-1)^{7-s}+2b_s(x),2(x-1)^s\rangle,$$
where $1\leq s\leq 7$ and
\begin{description}
\item{}
  $b_1(x)=b$ and $b\in \mathbb{F}_{2^m}$ arbitrary.

\item{}
  $b_2(x)=a_1+(a_1+a_2)(x-1)$ and $a_1,a_2\in\mathbb{F}_{2^m}$ arbitrary.

\item{}
  $b_3(x)=a_{3}(x-1)+a_{4}(x-1)^2$
 and $a_3,a_4\in\mathbb{F}_{2^m}$ arbitrary.

\item{}
$b_4(x)=a_3+a_5(x-1)^2+(a_3+a_5+a_6)(x-1)^3$
and $a_3,a_5,a_6\in\mathbb{F}_{2^m}$ arbitrary.

\item{}
$b_5(x)=a_5(x-1)+a_5(x-1)^2+(a_5+a_7)(x-1)^3+a_8(x-1)^4$
and $a_5,a_7,a_8\in\mathbb{F}_{2^m}$ arbitrary.

\item{}
$b_6(x)=a_5+a_5(x-1)+(a_5+a_7)(x-1)^2+a_9(x-1)^4+(a_9+a_{10})(x-1)^5$
and $a_5,a_7,a_9,a_{10}\in\mathbb{F}_{2^m}$ arbitrary.

\item{}
$b_7(x)=a_7(x-1)+a_9(x-1)^3+a_9(x-1)^4+(a_9+a_{11})(x-1)^5+(a_9+a_{12})(x-1)^6$
and $a_7,a_9,a_{11},a_{12}\in\mathbb{F}_{2^m}$ arbitrary.
\end{description}

\vskip 3mm \noindent
 {\bf Example 2}
 All $1+2^m+2(2^{2m}+2^{3m}+2^{4m}+2^{5m}+2^{6m}+2^{7m}+2^{8m})$
self-dual cyclic codes of length $32$ over $R$ are given by the following two cases:

\par
   (i) $1$ code: $\langle 2\rangle$.

\par
   (ii) $2^m+2(2^{2m}+2^{3m}+2^{4m}+2^{5m}+2^{6m}+2^{7m}+2^{8m})$ codes:
$$\langle (x-1)^{32-s}+2(x-1)^{15-s}+2b_s(x),2(x-1)^s\rangle,$$
where $1\leq s\leq 15$ and
\begin{description}
\item{}
  $b_j(x)$ is the same as that in Example 1, for all $j=1,2,3,4,5,6,7$;

\item{}
 $b_8(x)=a_7+a_9(x-1)^2+ a_9(x-1)^3+ (a_9 + a_{11})(x-1)^4+ a_9(x-1)^5+ (a_9 + a_{13})(x-1)^6 +(a_7 + a_9 + a_{11} + a_{13} + a_{14})(x-1)^7$;

\item{}
 $b_9(x)=a_9(x-1)+ a_9(x-1)^2+ (a_9 + a_{11})(x-1)^3 + a_9(x-1)^4+ (a_9 + a_{13})(x-1)^5 + (a_9 + a_{11} + a_{13})(x-1)^6+  (a_9 + a_{11} + a_{13} + a_{15})(x-1)^7 +a_{16}(x-1)^8$;

\item{}
 $b_{10}(x)=a_9 + a_9(x-1)+ (a_9 + a_{11})(x-1)^2+ a_9(x-1)^3+ (a_9 + a_{13})(x-1)^4+ (a_9 + a_{11} + a_{13})(x-1)^5
 + (a_9 + a_{11} + a_{13} + a_{15})(x-1)^6+ a_{17} (x-1)^8    +   (a_{17}  + a_{18})(x-1)^9 $;

\item{}
 $b_{11}(x)=a_{11}(x-1)+ a_{13}(x-1)^3+ (a_{11} + a_{13})(x-1)^4+ (a_{11} + a_{13} + a_{15})(x-1)^5+ a_{17}(x-1)^7+ a_{17}(x-1)^8
   + (a_{17} + a_{19})(x-1)^9     +  (a_{17} + a_{20} )(x-1)^{10} $;

\item{}
 $b_{12}(x)=a_{11}+ a_{13}(x-1)^2+ (a_{11} + a_{13})(x-1)^3 + (a_{11} + a_{13} + a_{15})(x-1)^4+ a_{17}(x-1)^6
 + a_{17}(x-1)^7 + (a_{17} + a_{19})(x-1)^8 + a_{17}(x-1)^9+ (a_{17} + a_{21})(x-1)^{10}
 +   (a_{17} + a_{19} + a_{21} + a_{22})(x-1)^{11} $;

\item{}
 $b_{13}(x)=a_{13}(x-1)+ a_{13}(x-1)^2+ (a_{13} + a_{15})(x-1)^3+ a_{17}(x-1)^5+ a_{17} (x-1)^6
 + (a_{17} + a_{19})(x-1)^7+ a_{17} (x-1)^8+ (a_{17} + a_{21})(x-1)^9 + (a_{17} + a_{19} + a_{21})(x-1)^{10}
 + (a_{17} + a_{19} + a_{21} + a_{23})(x-1)^{11}  +(a_{17} + a_{24})(x-1)^{12}$;

\item{}
 $b_{14}(x)=a_{13}+ a_{13}(x-1)+ (a_{13} + a_{15})(x-1)^2+ a_{17}(x-1)^4+ a_{17}(x-1)^5+ (a_{17} + a_{19})(x-1)^6
 + a_{17}(x-1)^7+ (a_{17} + a_{21})(x-1)^8+ (a_{17} + a_{19} + a_{21})(x-1)^9
 + (a_{17} + a_{19} + a_{21} + a_{23})(x-1)^{10}+ a_{17}(x-1)^{11}
 + (a_{17} + a_{25})(x-1)^{12}+(a_{17} + a_{19} + a_{25}  + a_{26})(x-1)^{13}$;

\item{}
 $b_{15}(x)=a_{15}(x-1)+ a_{17}(x-1)^3+ a_{17}(x-1)^4+ (a_{17} + a_{19} )(x-1)^5+ a_{17}(x-1)^6
 + (a_{17} + a_{21})(x-1)^7+ (a_{17} + a_{19} + a_{21})(x-1)^8+ (a_{17} + a_{19} + a_{21} + a_{23})(x-1)^9
 + a_{17}(x-1)^{10} + (a_{17} + a_{25})(x-1)^{11} + (a_{17} + a_{19} + a_{25})(x-1)^{12}
  +(a_{17} + a_{19} + a_{25} + a_{27})(x-1)^{13}  +(a_{17} + a_{21} + a_{25} + a_{28})(x-1)^{14}$,
\end{description}
\noindent
and $a_i\in\mathbb{F}_{2^m}$ arbitrary, for all integers $i$:
$1\leq i\leq 28$.

\vskip 3mm \noindent
 {\bf Example 3}
 All $1+2^m+2\sum_{t=2}^{16}(2^m)^t$
self-dual cyclic codes of length $64$ over $R$ are given by the following two cases:

\par
   (i) $1$ code: $\langle 2\rangle$.

\par
   (ii) $2^m+2\sum_{t=2}^{16}(2^m)^t $ codes:
$$\langle (x-1)^{64-s}+2(x-1)^{31-s}+2b_s(x),2(x-1)^s\rangle,$$
where $1\leq s\leq 31$ and
\begin{description}
\item{}
  $b_j(x)$ is the same as that in Examples 1 and 2, for all integers $j$: $1\leq j\leq 15$;

\item{}
 $b_{16}(x)=a_{15}+ a_{17}(x-1)^2+ a_{17}(x-1)^3+ (a_{17} + a_{19})(x-1)^4 + a_{17}(x-1)^5 + (a_{17} + a_{21})(x-1)^6 + (a_{17} + a_{19} + a_{21})(x-1)^7 + (a_{17} + a_{19} + a_{21} + a_{23})(x-1)^8 + a_{17}(x-1)^9  + (a_{17} + a_{25})(x-1)^{10} + (a_{17} + a_{19} + a_{25})(x-1)^{11}  + (a_{17} + a_{19} + a_{25} + a_{27})(x-1)^{12}  + (a_{17} + a_{21} + a_{25})(x-1)^{13}+ (a_{17} + a_{21} + a_{25} + a_{29})(x-1)^{14}+ (a_{15} + a_{17} + a_{19} + a_{21} + a_{23} + a_{25} + a_{27} + a_{29} + a_{30})(x-1)^{15}  $;

\item{}
 $b_{17}(x)= a_{17}(x-1)+ a_{17}(x-1)^2+ (a_{17} + a_{19})(x-1)^3  + a_{17}(x-1)^4 + (a_{17} + a_{21})(x-1)^5 + (a_{17} + a_{19} + a_{21})(x-1)^6 + (a_{17} + a_{19} + a_{21} + a_{23})(x-1)^7+ a_{17}(x-1)^8+ (a_{17} + a_{25})(x-1)^9  + (a_{17} + a_{19} + a_{25})(x-1)^{10} + (a_{17} + a_{19} + a_{25} + a_{27})(x-1)^{11} + (a_{17} + a_{21} + a_{25})(x-1)^{12}  + (a_{17} + a_{21} + a_{25} + a_{29})(x-1)^{13} + (a_{17} + a_{19} + a_{21} + a_{23} + a_{25} + a_{27} + a_{29})(x-1)^{14} + (a_{17} + a_{19} + a_{21} + a_{23} + a_{25} + a_{27} + a_{29} + a_{31})(x-1)^{15} +a_{32}(x-1)^{16} $;

\item{}
 $b_{18}(x)= a_{17}+ a_{17}(x-1) + (a_{17} + a_{19})(x-1)^2+ a_{17}(x-1)^3 + (a_{17} + a_{21})(x-1)^4 + (a_{17} + a_{19} + a_{21})(x-1)^5 + (a_{17} + a_{19} + a_{21} + a_{23})(x-1)^6 + a_{17}(x-1)^7+ (a_{17} + a_{25})(x-1)^8  + (a_{17} + a_{19} + a_{25})(x-1)^9 + (a_{17} + a_{19} + a_{25} + a_{27})(x-1)^{10}  + (a_{17} + a_{21} + a_{25})(x-1)^{11} + (a_{17} + a_{21} + a_{25} + a_{29})(x-1)^{12} + (a_{17} + a_{19} + a_{21} + a_{23} + a_{25} + a_{27} + a_{29})(x-1)^{13} + (a_{17} + a_{19} + a_{21} + a_{23} + a_{25} + a_{27} + a_{29} + a_{31})(x-1)^{14}+ a_{33}(x-1)^{16} +(a_{33} + a_{34})(x-1)^{17}  $;

\item{}
 $b_{19}(x)= a_{19}(x-1)+ a_{21}(x-1)^3 + (a_{19} + a_{21})(x-1)^4+ (a_{19} + a_{21} + a_{23})(x-1)^5+ a_{25}(x-1)^7  + (a_{19} + a_{25})(x-1)^8+ (a_{19} + a_{25} + a_{27})(x-1)^9  + (a_{21} + a_{25})(x-1)^{10} + (a_{21} + a_{25} + a_{29})(x-1)^{11} + (a_{19} + a_{21} + a_{23} + a_{25} + a_{27} + a_{29})(x-1)^{12} + (a_{19} + a_{21} + a_{23} + a_{25} + a_{27} + a_{29} + a_{31})(x-1)^{13} + a_{33}(x-1)^{15}  + a_{33}(x-1)^{16}+ (a_{33} + a_{35})(x-1)^{17}+(a_{33} + a_{36})(x-1)^{18}   $;

\item{}
 $b_{20}(x)=a_{19}+ a_{21}(x-1)^2+ (a_{19} + a_{21})(x-1)^3 + (a_{19} + a_{21} + a_{23})(x-1)^4+ a_{25}(x-1)^6 + (a_{19} + a_{25})(x-1)^7  + (a_{19} + a_{25} + a_{27})(x-1)^8 + (a_{21} + a_{25})(x-1)^9+ (a_{21} + a_{25} + a_{29})(x-1)^{10} + (a_{19} + a_{21} + a_{23} + a_{25} + a_{27} + a_{29})(x-1)^{11}  + (a_{19} + a_{21} + a_{23} + a_{25} + a_{27} + a_{29} + a_{31})(x-1)^{12} + a_{33}(x-1)^{14}  + a_{33}(x-1)^{15} + (a_{33} + a_{35})(x-1)^{16} + a_{33}(x-1)^{17}+ (a_{33} + a_{37})(x-1)^{18}  +
(a_{33} + a_{35} + a_{37} + a_{38})(x-1)^{19} $;

\item{}
 $b_{21}(x)=a_{21}(x-1) + a_{21}(x-1)^2 + (a_{21} + a_{23})(x-1)^3+ a_{25}(x-1)^5+ a_{25}(x-1)^6 + (a_{25} + a_{27})(x-1)^7+ (a_{21} + a_{25})(x-1)^8  + (a_{21} + a_{25} + a_{29})(x-1)^9+ (a_{21} + a_{23} + a_{25} + a_{27} + a_{29})(x-1)^{10}  + (a_{21} + a_{23} + a_{25} + a_{27} + a_{29} + a_{31})(x-1)^{11}+ a_{33}(x-1)^{13}+ a_{33}(x-1)^{14} + (a_{33} + a_{35})(x-1)^{15} + a_{33}(x-1)^{16} + (a_{33} + a_{37})(x-1)^{17} + (a_{33} + a_{35} + a_{37})(x-1)^{18} + (a_{33} + a_{35} + a_{37} + a_{39})(x-1)^{19} +(a_{33} + a_{40})(x-1)^{20} $;

\item{}
 $b_{22}(x)= a_{21}+ a_{21}(x-1)+ (a_{21} + a_{23})(x-1)^2+ a_{25}(x-1)^4+ a_{25}(x-1)^5+ (a_{25} + a_{27})(x-1)^6+ (a_{21} + a_{25})(x-1)^7+ (a_{21} + a_{25} + a_{29})(x-1)^8+ (a_{21} + a_{23} + a_{25} + a_{27} + a_{29})(x-1)^9+ (a_{21} + a_{23} + a_{25} + a_{27} + a_{29} + a_{31})(x-1)^{10}+ a_{33}(x-1)^{12}+ a_{33}(x-1)^{13} + (a_{33} + a_{35})(x-1)^{14}+ a_{33}(x-1)^{15}+ (a_{33} + a_{37})(x-1)^{16}+ (a_{33} + a_{35} + a_{37})(x-1)^{17}+ (a_{33} + a_{35} + a_{37} + a_{39})(x-1)^{18} + a_{33}(x-1)^{19}  + (a_{33} + a_{41})(x-1)^{20}   +(a_{33} + a_{35} + a_{41} + a_{42})(x-1)^{21} $;

\item{}
 $b_{23}(x)=a_{23}(x-1) + a_{25}(x-1)^3+ a_{25}(x-1)^4+ (a_{25} + a_{27})(x-1)^5+ a_{25}(x-1)^6 + (a_{25} + a_{29})(x-1)^7+ (a_{23} + a_{25} + a_{27} + a_{29})(x-1)^8+ (a_{23} + a_{25} + a_{27} + a_{29} + a_{31})(x-1)^9 + a_{33}(x-1)^{11}+ a_{33}(x-1)^{12}   + (a_{33} + a_{35})(x-1)^{13} + a_{33}(x-1)^{14}+ (a_{33} + a_{37})(x-1)^{15}+ (a_{33} + a_{35} + a_{37})(x-1)^{16} + (a_{33} + a_{35} + a_{37} + a_{39})(x-1)^{17} + a_{33}(x-1)^{18} + (a_{33} + a_{41})(x-1)^{19}+ (a_{33} + a_{35} + a_{41})(x-1)^{20}+ (a_{33} + a_{35} + a_{41} + a_{43})(x-1)^{21}   +    (a_{33} + a_{37} + a_{41} + a_{44})(x-1)^{22}   $;

\item{}
 $b_{24}(x)=a_{23}+ a_{25}(x-1)^2+ a_{25}(x-1)^3+ (a_{25} + a_{27})(x-1)^4 + a_{25}(x-1)^5 + (a_{25} + a_{29})(x-1)^6+ (a_{23} + a_{25} + a_{27} + a_{29})(x-1)^7+ (a_{23} + a_{25} + a_{27} + a_{29} + a_{31})(x-1)^8+ a_{33}(x-1)^{10} + a_{33}(x-1)^{11}+ (a_{33} + a_{35})(x-1)^{12}  + a_{33}(x-1)^{13} + (a_{33} + a_{37})(x-1)^{14} + (a_{33} + a_{35} + a_{37})(x-1)^{15}  + (a_{33} + a_{35} + a_{37} + a_{39})(x-1)^{16}+ a_{33}(x-1)^{17}   + (a_{33} + a_{41})(x-1)^{18}  + (a_{33} + a_{35} + a_{41})(x-1)^{19}  + (a_{33} + a_{35} + a_{41} + a_{43})(x-1)^{20}  + (a_{33} + a_{37} + a_{41})(x-1)^{21}  + (a_{33} + a_{37} + a_{41} + a_{45})(x-1)^{22} +(a_{33} + a_{35} + a_{37} + a_{39} + a_{41} + a_{43} + a_{45} + a_{46})(x-1)^{23} $;

\item{}
 $b_{25}(x)=a_{25}(x-1)+ a_{25}(x-1)^2 + (a_{25} + a_{27})(x-1)^3 + a_{25}(x-1)^4+ (a_{25} + a_{29})(x-1)^5+ (a_{25} + a_{27} + a_{29})(x-1)^6 + (a_{25} + a_{27} + a_{29} + a_{31})(x-1)^7 + a_{33}(x-1)^9 + a_{33}(x-1)^{10} + (a_{33} + a_{35})(x-1)^{11} + a_{33}(x-1)^{12} + (a_{33} + a_{37})(x-1)^{13} + (a_{33} + a_{35} + a_{37})(x-1)^{14}+ (a_{33} + a_{35} + a_{37} + a_{39})(x-1)^{15} + a_{33}(x-1)^{16}   + (a_{33} + a_{41})(x-1)^{17} + (a_{33} + a_{35} + a_{41})(x-1)^{18}+ (a_{33} + a_{35} + a_{41} + a_{43})(x-1)^{19}+ (a_{33} + a_{37} + a_{41})(x-1)^{20}+ (a_{33} + a_{37} + a_{41} + a_{45})(x-1)^{21}+ (a_{33} + a_{35} + a_{37} + a_{39} + a_{41} + a_{43} + a_{45})(x-1)^{22}   + (a_{33} + a_{35} + a_{37} + a_{39} + a_{41} + a_{43} + a_{45} + a_{47})(x-1)^{23}  + (a_{33} + a_{48})(x-1)^{24}   $;

\item{}
 $b_{26}(x)=a_{25}+ a_{25}(x-1)+ (a_{25} + a_{27})(x-1)^2 + a_{25}(x-1)^3  + (a_{25} + a_{29})(x-1)^4 + (a_{25} + a_{27} + a_{29})(x-1)^5+ (a_{25} + a_{27} + a_{29} + a_{31})(x-1)^6 + a_{33}(x-1)^8   + a_{33}(x-1)^9 + (a_{33} + a_{35})(x-1)^{10} + a_{33}(x-1)^{11} + (a_{33} + a_{37})(x-1)^{12} + (a_{33} + a_{35} + a_{37})(x-1)^{13}+ (a_{33} + a_{35} + a_{37} + a_{39})(x-1)^{14} + a_{33}(x-1)^{15}+ (a_{33} + a_{41})(x-1)^{16}   + (a_{33} + a_{35} + a_{41})(x-1)^{17} + (a_{33} + a_{35} + a_{41} + a_{43})(x-1)^{18}+ (a_{33} + a_{37} + a_{41})(x-1)^{19} + (a_{33} + a_{37} + a_{41} + a_{45})(x-1)^{20} + (a_{33} + a_{35} + a_{37} + a_{39} + a_{41} + a_{43} + a_{45})(x-1)^{21} + (a_{33} + a_{35} + a_{37} + a_{39} + a_{41} + a_{43} + a_{45} + a_{47})(x-1)^{22}+ a_{33}(x-1)^{23}  + (a_{33} + a_{49})(x-1)^{24}  +(a_{33} + a_{35} + a_{49} + a_{50})(x-1)^{25}  $;

\item{}
 $b_{27}(x)=a_{27}(x-1)+ a_{29}(x-1)^3+ (a_{27} + a_{29})(x-1)^4 + (a_{27} + a_{29} + a_{31})(x-1)^5+ a_{33}(x-1)^7  + a_{33}(x-1)^8  + (a_{33} + a_{35})(x-1)^9 + a_{33}(x-1)^{10} + (a_{33} + a_{37})(x-1)^{11} + (a_{33} + a_{35} + a_{37})(x-1)^{12} + (a_{33} + a_{35} + a_{37} + a_{39})(x-1)^{13} + a_{33}(x-1)^{14} + (a_{33} + a_{41})(x-1)^{15} + (a_{33} + a_{35} + a_{41})(x-1)^{16}+ (a_{33} + a_{35} + a_{41} + a_{43})(x-1)^{17}  + (a_{33} + a_{37} + a_{41})(x-1)^{18} + (a_{33} + a_{37} + a_{41} + a_{45})(x-1)^{19}  + (a_{33} + a_{35} + a_{37} + a_{39} + a_{41} + a_{43} + a_{45})(x-1)^{20} + (a_{33} + a_{35} + a_{37} + a_{39} + a_{41} + a_{43} + a_{45} + a_{47})(x-1)^{21}+ a_{33}(x-1)^{22}+ (a_{33} + a_{49})(x-1)^{23}  + (a_{33} + a_{35} + a_{49})(x-1)^{24} + (a_{33} + a_{35} + a_{49} + a_{51})(x-1)^{25} + (a_{33} + a_{37} + a_{49} + a_{52})(x-1)^{26}
 $;

\item{}
 $b_{28}(x)=a_{27}+ a_{29}(x-1)^2 + (a_{27} + a_{29})(x-1)^3 + (a_{27} + a_{29} + a_{31})(x-1)^4+ a_{33}(x-1)^6+ a_{33}(x-1)^7+ (a_{33} + a_{35})(x-1)^8+ a_{33}(x-1)^9 + (a_{33} + a_{37})(x-1)^{10}  + (a_{33} + a_{35} + a_{37})(x-1)^{11}+ (a_{33} + a_{35} + a_{37} + a_{39})(x-1)^{12}   + a_{33}(x-1)^{13}+ (a_{33} + a_{41})(x-1)^{14} + (a_{33} + a_{35} + a_{41})(x-1)^{15} + (a_{33} + a_{35} + a_{41} + a_{43})(x-1)^{16}  + (a_{33} + a_{37} + a_{41})(x-1)^{17} + (a_{33} + a_{37} + a_{41} + a_{45})(x-1)^{18} + (a_{33} + a_{35} + a_{37} + a_{39} + a_{41} + a_{43} + a_{45})(x-1)^{19}+ (a_{33} + a_{35} + a_{37} + a_{39} + a_{41} + a_{43} + a_{45} + a_{47})(x-1)^{20} + a_{33}(x-1)^{21} + (a_{33} + a_{49})(x-1)^{22} + (a_{33} + a_{35} + a_{49})(x-1)^{23} + (a_{33} + a_{35} + a_{49} + a_{51})(x-1)^{24}+ (a_{33} + a_{37} + a_{49})(x-1)^{25}+ (a_{33} + a_{37} + a_{49} + a_{53})(x-1)^{26}+(a_{33} + a_{35} + a_{37} + a_{39} + a_{49} + a_{51} + a_{53} + a_{54})(x-1)^{27}  $;

\item{}
 $b_{29}(x)= a_{29}(x-1)+ a_{29}(x-1)^2 + (a_{29} + a_{31})(x-1)^3 + a_{33}(x-1)^5+ a_{33}(x-1)^6+ (a_{33} + a_{35})(x-1)^7 + a_{33}(x-1)^8 + (a_{33} + a_{37})(x-1)^9+ (a_{33} + a_{35} + a_{37})(x-1)^{10}  + (a_{33} + a_{35} + a_{37} + a_{39})(x-1)^{11}  + a_{33}(x-1)^{12} + (a_{33} + a_{41})(x-1)^{13}+ (a_{33} + a_{35} + a_{41})(x-1)^{14}   + (a_{33} + a_{35} + a_{41} + a_{43})(x-1)^{15} + (a_{33} + a_{37} + a_{41})(x-1)^{16}  + (a_{33} + a_{37} + a_{41} + a_{45})(x-1)^{17} + (a_{33} + a_{35} + a_{37} + a_{39} + a_{41} + a_{43} + a_{45})(x-1)^{18}+ (a_{33} + a_{35} + a_{37} + a_{39} + a_{41} + a_{43} + a_{45} + a_{47})(x-1)^{19} + a_{33}(x-1)^{20} + (a_{33} + a_{49})(x-1)^{21}  + (a_{33} + a_{35} + a_{49})(x-1)^{22} + (a_{33} + a_{35} + a_{49} + a_{51})(x-1)^{23}+ (a_{33} + a_{37} + a_{49})(x-1)^{24}  + (a_{33} + a_{37} + a_{49} + a_{53})(x-1)^{25} + (a_{33} + a_{35} + a_{37} + a_{39} + a_{49} + a_{51} + a_{53})(x-1)^{26}+ (a_{33} + a_{35} + a_{37} + a_{39} + a_{49} + a_{51} + a_{53} + a_{55})(x-1)^{27}+(a_{33} + a_{41} + a_{49} + a_{56})(x-1)^{28} $;

\item{}
 $b_{30}(x)= a_{29}+ a_{29}(x-1) + (a_{29} + a_{31})(x-1)^2  + a_{33}(x-1)^4+ a_{33}(x-1)^5 + (a_{33} + a_{35})(x-1)^6 + a_{33}(x-1)^7+ (a_{33} + a_{37})(x-1)^8 + (a_{33} + a_{35} + a_{37})(x-1)^9 + (a_{33} + a_{35} + a_{37} + a_{39})(x-1)^{10}+ a_{33}(x-1)^{11} + (a_{33} + a_{41})(x-1)^{12}  + (a_{33} + a_{35} + a_{41})(x-1)^{13}+ (a_{33} + a_{35} + a_{41} + a_{43})(x-1)^{14} + (a_{33} + a_{37} + a_{41})(x-1)^{15} + (a_{33} + a_{37} + a_{41} + a_{45})(x-1)^{16}+ (a_{33} + a_{35} + a_{37} + a_{39} + a_{41} + a_{43} + a_{45})(x-1)^{17} + (a_{33} + a_{35} + a_{37} + a_{39} + a_{41} + a_{43} + a_{45} + a_{47})(x-1)^{18}+ a_{33}(x-1)^{19}    + (a_{33} + a_{49})(x-1)^{20} + (a_{33} + a_{35} + a_{49})(x-1)^{21} + (a_{33} + a_{35} + a_{49} + a_{51})(x-1)^{22} + (a_{33} + a_{37} + a_{49})(x-1)^{23}+ (a_{33} + a_{37} + a_{49} + a_{53})(x-1)^{24}+ (a_{33} + a_{35} + a_{37} + a_{39} + a_{49} + a_{51} + a_{53})(x-1)^{25} + (a_{33} + a_{35} + a_{37} + a_{39} + a_{49} + a_{51} + a_{53} + a_{55})(x-1)^{26} + (a_{33} + a_{41} + a_{49})(x-1)^{27} + (a_{33} + a_{41} + a_{49} + a_{57})(x-1)^{28}   +(a_{33} + a_{35} + a_{41} + a_{43} + a_{49} + a_{51} + a_{57} + a_{58})(x-1)^{29}  $;

\item{}
 $b_{31}(x)=a_{31}(x-1)+ a_{33}(x-1)^4 + (a_{33} + a_{35})(x-1)^5+ a_{33}(x-1)^6 + (a_{33} + a_{37})(x-1)^7  + (a_{33} + a_{35} + a_{37})(x-1)^8+ (a_{33} + a_{35} + a_{37} + a_{39})(x-1)^9 + a_{33}(x-1)^{10}  + (a_{33} + a_{41})(x-1)^{11} + (a_{33} + a_{35} + a_{41})(x-1)^{12} + (a_{33} + a_{35} + a_{41} + a_{43})(x-1)^{13} + (a_{33} + a_{37} + a_{41})(x-1)^{14} + (a_{33} + a_{37} + a_{41} + a_{45})(x-1)^{15} + (a_{33} + a_{35} + a_{37} + a_{39} + a_{41} + a_{43} + a_{45})(x-1)^{16} + (a_{33} + a_{35} + a_{37} + a_{39} + a_{41} + a_{43} + a_{45} + a_{47})(x-1)^{17} + a_{33}(x-1)^{18}  + (a_{33} + a_{49})(x-1)^{19} + (a_{33} + a_{35} + a_{49})(x-1)^{20}+ (a_{33} + a_{35} + a_{49} + a_{51})(x-1)^{21} + (a_{33} + a_{37} + a_{49})(x-1)^{22} + (a_{33} + a_{37} + a_{49} + a_{53})(x-1)^{23} + (a_{33} + a_{35} + a_{37} + a_{39} + a_{49} + a_{51} + a_{53})(x-1)^{24} + (a_{33} + a_{35} + a_{37} + a_{39} + a_{49} + a_{51} + a_{53} + a_{55})(x-1)^{25}+ (a_{33} + a_{41} + a_{49})(x-1)^{26}+ (a_{33} + a_{41} + a_{49} + a_{57})(x-1)^{27}  + (a_{33} + a_{35} + a_{41} + a_{43} + a_{49} + a_{51} + a_{57})(x-1)^{28}  + (a_{33} + a_{35} + a_{41} + a_{43} + a_{49} + a_{51} + a_{57} + a_{59})(x-1)^{29} +
(a_{33} + a_{37} + a_{41} + a_{45} + a_{49} + a_{53} + a_{57} + a_{60})(x-1)^{30} $,
\end{description}

\noindent
and $a_i\in\mathbb{F}_{2^m}$ arbitrary, for all integers $i$:
$1\leq i\leq 60$.

\vskip 3mm
\noindent
  {\bf Remarks} (i) Let $m=1$. Then $\mathbb{F}_2=\{0,1\}$ and we have constructed
the following self-dual cyclic codes over $\mathbb{Z}_4$:
\begin{description}
\item{}
  $59$ self-dual cyclic codes of length $16$ given by Example 1.

\item{}
  $1019$ self-dual cyclic codes of length $32$ given by Example 2.
\item{}
  $262139$ self-dual cyclic codes of length $64$ given by Example 3.
\end{description}

\par
   (ii) Let $m=2$ and set
${\rm GR}(4,2)=\frac{\mathbb{Z}_4[z]}{\langle z^2+z+1\rangle}=\{a+bz\mid a,b\in \mathbb{Z}_4\}$
in which the arithmetic is done modulo $z^2+z+1$.  Then
$$\mathbb{F}_4=\{c+dz\mid c,d\in \mathbb{F}_2\}=\{0,1,z,1+z\}$$
and we have constructed
the following self-dual cyclic codes over the Galois ring ${\rm GR}(4,2)$:
\begin{description}
\item{}
  $677$ self-dual cyclic codes of length $16$ given by Example 1.

\item{}
  $174757$ self-dual cyclic codes of length $32$ given by Example 2.
\item{}
  $11453246117$ self-dual cyclic codes of length $64$ given by Example 3.
\end{description}

\section{Conclusions}
\noindent
For any positive integers $m$ and $k$, we have given a simple and efficient method to construct all distinct Euclidean self-dual cyclic codes of length $2^k$ over the Galois ring ${\rm GR}(4,m)$, using Kronecker products of matrices over $\mathbb{Z}_2$ of specific types.
On this basis,
by use of combination numbers,
we provide an explicit expression to accurately represent this class of Euclidean self-dual cyclic codes.
Based on these explicit expressions, we can further study the parametric properties of this class of  self-dual cyclic codes.



\vskip 3mm \noindent {\bf Acknowledgments}
Part of this work was done when Yonglin Cao was visiting Chern Institute of Mathematics,
Nankai University, Tianjin, China. He thanks the institution for the kind hospitality. This research is
supported in part by the National Natural Science Foundation of
China (Grant Nos. 11671235, 11801324),  the Shandong Provincial Natural Science Foundation, China
(Grant No. ZR2018BA007), the Nanyang Technological University Research (Grant No. M4080456), the Scientific Research Fund of Hubei Provincial Key Laboratory of Applied Mathematics (Hubei University)(Grant Nos. HBAM201906, HBAM201804), the Scientific Research Fund of Hunan Provincial Key Laboratory of Mathematical Modeling and Analysis in Engineering (No. 2018MMAEZD09) and the Visiting Project Funds of Shandong University of Technology (SDUT).


\end{document}